\documentclass[aps,prx,amsmath,amssymb,footinbib,showpacs,twocolumn,superscriptaddress]{revtex4-1}
\usepackage{amsmath}
\usepackage{amssymb}
\usepackage{amsthm}
\usepackage{setspace}
\usepackage{graphicx}
\usepackage{braket}
\usepackage{mathrsfs}
\usepackage{float}
\usepackage[colorlinks = true,linkcolor = red,urlcolor  = blue,citecolor = blue,anchorcolor = blue]{hyperref}

\newcommand{\nn}{\nonumber \\}
\newcommand{\angstrom}{\text{\normalfont\AA}}
\newcommand{\ua}{\uparrow}
\newcommand{\da}{\downarrow}

\begin{document}


\title{Conductance feature smearing and anisotropic suppression of induced superconductivity in a Majorana nanowire}

\author{Chun-Xiao Liu}
\affiliation{
Condensed Matter Theory Center and Joint Quantum Institute, Department of Physics, University of Maryland, College Park, MD 20742, USA}

\author{Jay D. Sau}
\affiliation{
Condensed Matter Theory Center and Joint Quantum Institute, Department of Physics, University of Maryland, College Park, MD 20742, USA}

\author{Tudor D. Stanescu}
\affiliation{
Condensed Matter Theory Center and Joint Quantum Institute, Department of Physics, University of Maryland, College Park, MD 20742, USA}
\affiliation{
Department of Physics and Astronomy, West Virginia University, Morgantown, WV 26506}

\author{S. Das Sarma}
\affiliation{
Condensed Matter Theory Center and Joint Quantum Institute, Department of Physics, University of Maryland, College Park, MD 20742, USA}

\date{\today}

\begin{abstract}
In a recent high-quality experimental work on normal metal--superconducting nanowire junctions [J.D.S. Bommer \textit{et al}., \href{https://arxiv.org/abs/1807.01940}{arXiv:1807.01940} (2018)], strong anisotropic suppression of induced superconductivity has been observed in tunnel conductance measurements in the presence of applied magnetic field with variable orientation. Following this finding, we investigate theoretically the dependence of tunnel conductance on the direction of the Zeeman field in order to understand the operational mechanisms and to extract effective system parameters. Second, motivated by a generic discrepancy between experiment and theory, i.e., many in-gap and above-gap conductance features predicted by theory are barely observed in experiments, we study several mechanisms possibly responsible for the suppression of the theoretically predicted conductance features (e.g., length of the nanowire, self-energy effect due to the proximity effect, finite temperature, finite dissipation, and multiband effect). One essential finding in the current work is that only by a combined understanding of both suppression mechanisms can we extract effective system parameters from the experimental data (e.g., the effective nanowire-superconductor coupling, the effective Land\'e $g$ factor, and the chemical potential of the semiconducting nanowire). In addition, we consider topologically trivial Andreev bound states in hybrid nanowires in the presence of potential inhomogeneities, such as external quantum dots or potential inhomogeneities inside the nanowire. We compare the anisotropic, field-dependent features induced by these nontopological Andreev bound states with the corresponding features produced by topological Majorana zero modes in pristine nanostructures, so that we can provide guidance to differentiate between the topologically trivial and nontrivial cases.
\end{abstract}

\maketitle

\section{introduction}

Topological superconductivity (TSC) hosting exotic zero-energy Majorana excitations has become an area of  intense research, both theoretically and experimentally, over the past decade~\cite{Nayak2008Non-Abelian, Alicea2012New, Leijnse2012Introduction, Beenakker2013Search, Stanescu2013Majorana, Jiang2013Non, DasSarma2015Majorana, Elliott2015Colloquium, Sato2016Majorana, Sato2017Topological, Aguado2017Majorana, Lutchyn2018Majorana}. The most-studied realization of one-dimensional TSC is a strongly spin-orbit-coupled (SOC) semiconducting nanowire in proximity to a conventional $s$-wave superconductor (SC) and in the presence of an external Zeeman field~\cite{Sau2010Generic, Lutchyn2010Majorana, Oreg2010Helical, Sau2010NonAbelian}. When the Zeeman field applied along the wire axis exceeds a certain critical value, the hybrid nanowire becomes an effectively spinless $p$-wave SC that hosts a pair of zero-energy Majorana zero modes (MZM) localized at the opposite ends of the nanowire~\cite{Kitaev2001Unpaired}. These spatially separated MZMs are robust against local perturbations and obey non-Abelian statistics as long as their separation is large enough~\cite{Ivanov2001NonAbelian, Alicea2011NonAbelian}. Based on these predicted properties, hybrid semiconductor-superconductor (SM-SC) nanowires have become the most promising platform for fault-tolerant topological quantum computation, and are being actively studied in many laboratories both for their intriguing fundamental topological properties and their technological prospects in creating a commercial quantum computer.~\cite{Aasen2016Milestones, Vijay2016TeleportationBased, Plugge2017Majorana, Karzig2017Scalable}.

The most direct method for detecting the presence of MZMs in a hybrid nanowire is to measure the differential conductance for charge tunneling from a normal-metal lead into the end of the proximitized nanowire. The hallmark associated with the presence of MZMs is a zero-bias conductance peak (ZBCP) that has a quantized height of $2e^2/h$ at zero temperature arising from the perfect Andreev reflection by the zero-energy MZM~\cite{Sengupta2001Midgap, Law2009Majorana}. Sustained efforts over the past decade have generated significant improvements in materials science, nanofabrication, and measurements~\cite{Mourik2012Signatures, Das2012Zero, Deng2012Anomalous,Churchill2013Superconductor, Finck2013Anomalous, Albrecht2016Exponential, Chen2017Experimental, Deng2016Majorana, Zhang2017Ballistic, Gul2018Ballistic, Nichele2017Scaling, Zhang2017Quantized} and have led to the observation of stable zero-energy tunneling conductance signatures that are consistent with the presence of MZM in SM-SC hybrid structures. A major recent advance is the observation of a quantized tunnel conductance plateau that is to some extent stable against variations of different control parameters~\cite{Zhang2017Quantized}, as expected for MZM-induced ZBCPs. 

In spite of this remarkable experimental progress, there are a few crucial aspects that have not yet been elucidated, which, ultimately, makes the search for Majorana zero modes in solid state systems incomplete. The main concern is the possibility that, rather than being induced by MZMs localized at the ends of the system, the characteristic experimental ZBCP signatures are actually generated by trivial Andreev bound states (ABSs) emerging near zero energy in systems with a position-dependent effective potential~\cite{Kells2012Near, Prada2012Transport, Liu2017Andreev,Chiu2017Conductance, Setiawan2017Electron,Moore2018a,Liu2018Distinguishing,Fleckenstein2018Decaying, Moore2018b,Vuik2018Reproducing}. In addition, there are a few significant discrepancies between the theoretical predictions and the experimental tunneling conductance data. For example, theory predicts~\cite{Sau2010NonAbelian, Stanescu2011Majorana} that the emergence of MZM should be accompanied by the closing and reopening of the bulk quasiparticle gap at the topological quantum phase transition (TQPT), as the TSC transitions from the trivial to the topological phase accompanied by the bulk gap closure. However, clear, unambiguous experimental signatures associated with the closing and reopening of the {\em bulk} gap have been rarely (if ever) observed. Moreover,  numerical simulations of normal-metal lead--SC nanowire systems predict a rich structure of the differential conductance as function of the applied Zeeman field~\cite{Liu2017Role}, yet the experimental observations reveal much simpler smooth behavior with no detailed fine structures. This not only puts into question the accuracy of the theoretical modeling, but deprives one of potentially valuable information that could help with adjusting the model parameters and, ultimately, optimizing the hybrid nanowire devices for the decisive manifestation of MZMs. 

Clarifying these outstanding issues involves, in addition to exploring new experimental setups~\cite{Chiu2017Interference, Chiu2018Fractional, Liu2018Measuring}, overcoming at least two key challenges: (i) understanding in detail the mechanisms responsible for the suppression of certain predicted spectral features in charge tunneling experiments, and (ii) optimizing the techniques for extracting effective system parameters from the experimental data. To address these challenges, it is important to expand the comparison between theory and experiment over a parameter subspace as large as possible. In the current work, we try to achieve both of these goals using a recent experimental publication~\cite{Bommer2018SpinOrbit} as the guide in order to bring theory and experiment closer to each other in their details.

An important degree of freedom characterizing the hybrid SM-SC parameter space is the angle between the applied magnetic field and the axis of the nanowire. So far, most of the research, in both theory and experiment, was concerned with  external Zeeman fields applied along the axis of the nanowire, with only the strength of the Zeeman field being varied. This is the field configuration where the localized MZMs are predicted to be the most prominent as the applied magnetic field increases. Recently, a high-quality experiment~\cite{Bommer2018SpinOrbit} has characterized in detail the dependence of the tunnel conductance on the direction of the applied magnetic field. First, the tunnel conductance was measured as a function of the strength of the applied field, with the direction of the field being fixed along the x-, y-, or z-direction (with the wire axis fixed along the x-direction). It was shown that the critical strength of the magnetic field associated with the closure of the superconducting gap has a strong dependence on the direction of the applied field. Second, the tunnel conductance was measured as a continuous function of the direction of the magnetic field in the plane perpendicular to the nanowire axis, with the strength of the field being fixed. The results reveal a strongly anisotropic suppression of the induced superconducting gap as function of the orientation of the applied magnetic field. These are some of the features we want to theoretically understand in depth in the current work.

In Ref.~\cite{Bommer2018SpinOrbit}, there are also numerical simulations for understanding the anisotropic suppression of superconductivity in the Majorana nanowire in the presence of applied magnetic field. But the simulations only explore a small volume in the parameter space and do not consider any inhomogeneity-induced topologically trivial bound states in the Majorana nanowire. In this work we address some of the outstanding problems facing the realization of MZM in SM-SC hybrid structures by investigating the mechanisms that control the low-energy spectroscopic features characterizing tunnel conductance measurements in the presence of applied magnetic fields with variable orientation. More specifically, we consider the following question: can one discriminate between MZM-induced topological features and the non-topological features produced by the topologically trivial ABS based on the dependence of gap closing signatures on the orientation of the applied magnetic field? Of course, to properly answer this question one has to also understand the mechanisms responsible for the suppression of certain spectral features in the tunneling conductance data. This effort of understanding the relationship between ``visible'' spectral features and the basic properties of the hybrid system represents an important step toward optimizing the techniques for extracting effective system parameters from the experimental data. If the well-accepted theory for MZMs in nanowires disagrees with experimental data in some qualitative manner (e.g., absence of fine conductance structures in the data), we must strive to understand this discrepancy by taking into account physical mechanisms typically not included in the minimal theory. In this work, we present three closely interconnected aspects of this effort. First, we investigate the rather generic absence of detailed in-gap and above-gap conductance features in the measured differential conductance maps. We consider several mechanisms possibly responsible for this behavior: the dependence of the spectral features on the length of the nanowire, self-energy effect due to the proximity to the parent superconductor, finite temperature, finite dissipation, and multi-band effect. Second, we study the dependence of the tunnel conductance on the direction of the Zeeman field in pristine nanowires. Inspired by the recent experiment~\cite{Bommer2018SpinOrbit}, we examine both the dependence on the strength of the applied field for different field orientations along the x-, y-, and z-axes and the continuous dependence on the field direction for a fixed field strength. Note that the dependence of the conductance on the strength of the applied field for different field orientations is studied within the context of our investigation of the mechanisms responsible for the suppression of fine conductance features in pristine nanowires. Third, we consider hybrid nanowires in the presence of potential inhomogeneities, such as external quantum dots or effective potential inhomogeneities inside the nanowire, and compare the anisotropic field-dependent features induced by the topologically trivial ABS emerging in these systems with the corresponding features produced by MZM in pristine nanostructures. Since the possible existence of trivial low-energy ABS mimicking MZM properties cannot \textit{a priori} be ruled out, it is important to investigate how the ABS behavior is modified   (as compared with the MZM behavior) due to the tuning of the external field direction.

The remainder of this paper is organized as follows. In Sec.~\ref{sec:theory}, we introduce the theoretical model that describes the normal metal--superconducting nanowire system, along with the numerical method for calculating the differential tunnel conductance. Section~\ref{sec:suppression} is devoted to studying the dependence of the tunnel conductance on the strength of the applied magnetic field for different field orientations, focusing on pristine Majorana nanowires. This is done within the context of our investigation of the mechanisms responsible for the suppression of differential conductance features. We also consider the continuous dependence of the tunnel conductance on the direction of the Zeeman field in the context of pristine Majorana nanowires. In Sec.~\ref{sec:ABS}, we consider hybrid nanowires in the presence of effective potential inhomogeneities giving rise to low-energy ABSs and study the suppression of the conductance features, as well as dependence on the field orientation. We compare these ABS results with the corresponding MZM results for pristine nanowires. Finally, in Sec.~\ref{sec:conclusion} we summarize the results and present our conclusions.

\section{Theoretical model and numerical method} \label{sec:theory}

\begin{figure}[tbp]
\centering
\includegraphics[width=1.0\columnwidth]{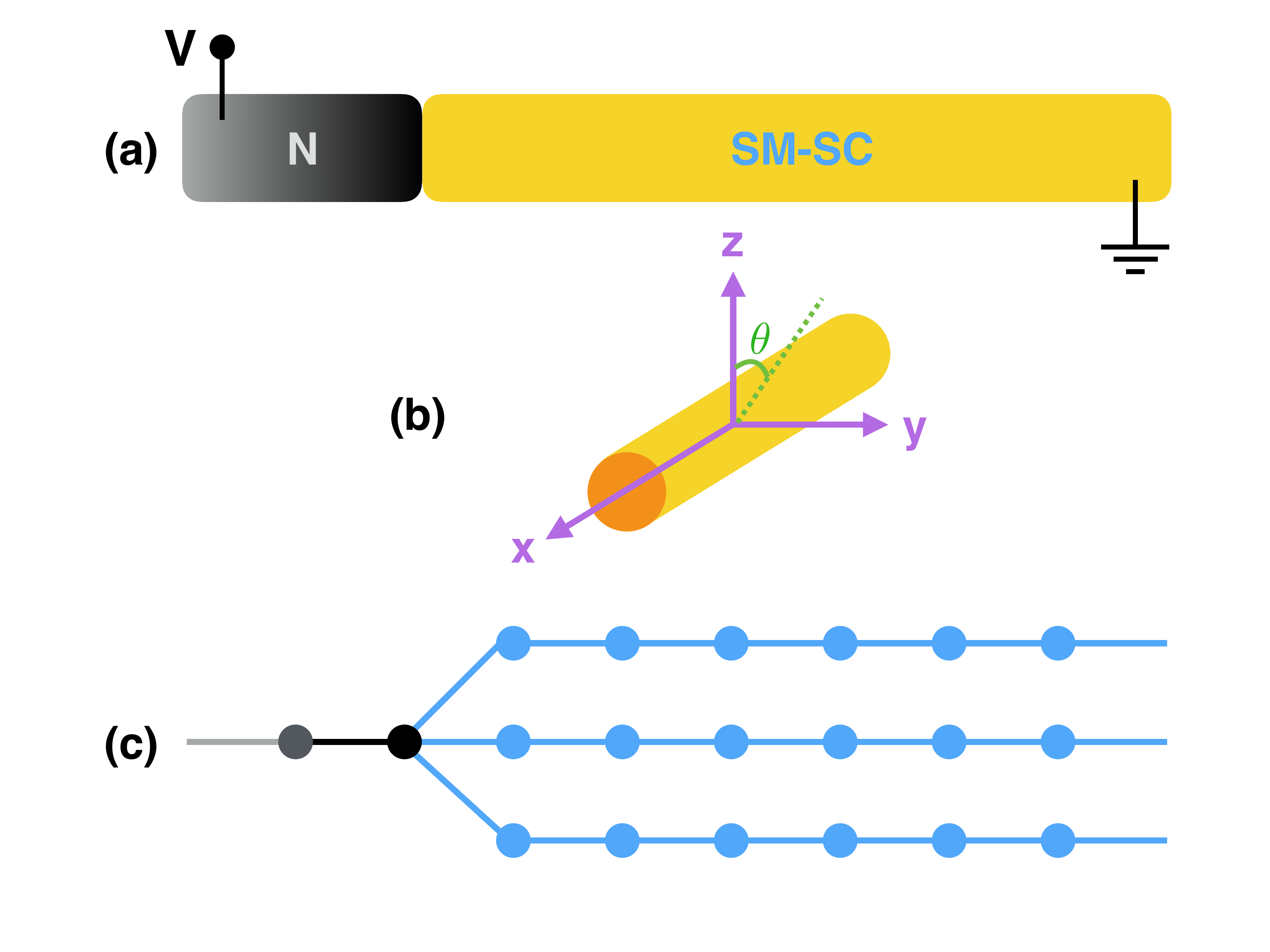}
\caption{ (a) Schematic representation of an NS junction in the SM-SC nanowire system. A normal-metal lead (black)  is in contact with an SM-SC nanowire (yellow). A bias voltage $V$ is applied on the normal-metal lead, while the SC is grounded. (b) Definition of the coordinate axes. The x-axis is along the direction of the nanowire, while the y- and z-axes span the plane perpendicular to the wire, with the y-axis parallel to the plane of the substrate and the z-axis normal to the substrate.  The angle $\theta$ (in the y-z plane) is measured from the z-axis. (c) Schematic representation of the NS junction with multiple subbands (in the lattice model). The normal-metal lead (black) always contains a single spinful band. The subbands from the SM-SC nanowire (blue) couple with the lead separately. How many transverse subbands are occupied depends on the details of the system and may not be known in general for specific experimental samples.}
\label{fig:schematic}
\end{figure}

A minimal effective theory for describing  the low-energy physics of a pristine one-dimensional semiconductor-superconductor nanowire (oriented along the x-axis) is given by 
\begin{align}
&\hat{H} = \frac{1}{2} \int^{L_{\text{wire}}}_0 dx \hat{\Psi}^{\dagger}(x) h_{\text{BdG}}  \hat{\Psi}(x) \nn
& h_{\text{BdG}} = h_{\text{nw}} + h_{\text{sc}} \nn
&h_{\text{nw}} = \left( -\frac{\partial^2_x}{2m^*} -i\alpha_R \partial_x \sigma_y - \mu \right)\tau_z + \vec{V}_Z \cdot \vec{\sigma} - i\Gamma \nn
&h_{\text{sc}} = \Delta_{\rm ind} \tau_x \quad  \text{or}  \quad h_{\text{sc}} = \Sigma(\omega) = -\lambda \frac{\omega \tau_0 + \Delta \tau_x}{\sqrt{\Delta^2-\omega^2}},
\label{eq:1d_smsc}
\end{align}
where we have taken $\hbar=1$,  $\hat{\Psi} = \Big( \hat{c}_{\ua}, \hat{c}_{\da}, \hat{c}^{\dagger}_{\da}, -\hat{c}^{\dagger}_{\ua} \Big)^{T}$ is a  Nambu spinor, and $h_{\text{BdG}}$ is the corresponding Bogoliubov-de Gennes (BdG) Hamiltonian. Here, $m^*$ is the effective mass, $\alpha_R$ the strength of the Rashba spin-orbit coupling, $\mu$ the chemical potential measured relative to the spin-orbit crossing point, $\vec{V}_Z = (V_{Zx},V_{Zy},V_{Zz} )$ the vector Zeeman field, which can point in any direction in the three-dimensional space, $\Gamma$ the amount of dissipation inside the nanowire (which is treated phenomenologically in the current work through the $i\Gamma$ term in the Hamiltonian), and $L_{\text{wire}}$ the length of the nanowire. The quantities $\sigma_{x,y,z}$ ($\tau_{x,y,z}$) are Pauli matrices acting on the spin (particle-hole) space. The term  $h_{\text{sc}}$, which captures the SC proximity effect, is considered within two different approximations: in the weak coupling limit (the usual approximation used in most nanowire calculations), $h_{\text{sc}}$ is approximated as an induced pairing term $\Delta_{\rm ind} \tau_x$, while, more generally, the proximity effect  is incorporated as the self-energy contribution $\Sigma(\omega)$, with $\lambda$ being the effective coupling strength between the SM wire and the parent SC. The inclusion of the self-energy term enables us to go beyond the weak-coupling proximity effect and consider situations where the SM-SC coupling is strong. The Hamiltonian for the normal-metal lead only contains the first three terms in $h_{\text{nw}}$, i.e., we do not include Zeeman effect, dissipation, and superconductivity. The chemical potential of the lead is set in the middle of the band, away from any Van Hove singularity. Note that in the current work, we only include a single spinful mode inside the normal-metal lead. In the context of quantum point contact, which is the case for Majorana nanowire tunnel spectroscopy experiments, although there are many modes in the metallic lead, the number of working mode in the lead is constrained by the number of transverse modes in the point contact area~\cite{Beenakker1992Quantum}. So effectively we can assume that there is only a single spinful mode in the metallic lead, being consistent with experimental observations. The parameter values used in the numerical calculations, motivated by the experimental SM-SC hybrid structures, are $m^*=0.015m_e$, $\alpha_R=0.5~$eV$\angstrom$, $\Gamma=0.008~$meV,  unless stated otherwise. Note that Eq.~\eqref{eq:1d_smsc} is a one-dimensional model of the nanowire that does not include orbital effects. However, in some instances, we will consider confinement-induced multiband effects (arising from the quantization of the transverse motion of the carriers in the nanowire) by combining several one-dimensional SM-SC nanowires, with possibly different wire parameters. In this work, we assume that there is no coupling between the subbands. Note that, regardless of the  number of occupied subbands in the SM wire, the normal-metal lead is assumed to contain a single spinful subband. A schematic representation of the normal metal--superconductor (NS) junction is shown in Fig.~\ref{fig:schematic}. Note that the model defined by Eq.~\eqref{eq:1d_smsc} is by definition for a pristine nanowire without any external inhomogeneous potential, and as such, does not have any extrinsic ABS mimicking MZM behavior. We defer our discussion of extrinsic ABS to Sec.~\ref{sec:ABS} where Eq.~\eqref{eq:1d_smsc} will be appropriately modified by the addition of an inhomogeneous potential term.

 The formula for the conductance through the NS junction is \begin{align}
G(V) = \frac{e^2}{h} \Big( N + {\rm Tr}\Big[ S^{\dagger}_{he}S_{he}(V) \Big]-{\rm Tr}\Big[ S^{\dagger}_{ee}S_{ee}(V) \Big] \Big),
\end{align} 
where $N$ is the number of spin polarized channels in the lead, and $S_{he}(S_{ee})$ is the S-matrix for an incoming electron being reflected as a hole (electron). In the current work, we consider a single spinful subband in the normal-metal lead, i.e., $N=2$, and thus the range of the conductance value is $0 < G < 4e^2/h$~\cite{Blonder1982Transition}. To numerically calculate the S-matrix, we first discretize the continuum Hamiltonian in Eq.~\eqref{eq:1d_smsc} into a corresponding tight-binding model. We then apply the Kwant numerical package~\cite{kwant} to obtain the S-matrix and the differential conductance. Note that even when the BdG Hamiltonian becomes non-Hermitian in the presence of dissipation, the calculation of the S matrix by the Kwant package is still feasible and reliable. 

\section{Suppression of fine conductance structures in pristine systems with different field orientations} \label{sec:suppression}

A generic characteristic of the tunnel spectroscopy of Majorana nanowires is that experimentally measured spectra~\cite{Mourik2012Signatures, Das2012Zero, Deng2012Anomalous,Churchill2013Superconductor, Finck2013Anomalous, Albrecht2016Exponential, Chen2017Experimental, Deng2016Majorana, Zhang2017Ballistic, Gul2018Ballistic, Nichele2017Scaling, Zhang2017Quantized} always show much less structure than numerical  conductance calculations~\cite{Lin2012Zero, Liu2017Role, Liu2017Andreev, Liu2017Phenomenology, Setiawan2017Electron}. For example, in the numerical simulations, one can typically see conductance features associated with the (bulk) gap closing and reopening at TQPT. Also,  one can clearly distinguish a nontrivial conductance structure above the induced SC gap. By contrast, in experiments the gap reopening feature is rather elusive, while the conductance above the SC gap is always structureless. These characteristics are rather generic in Majorana nanowire experiments (i.e., independent of which group is reporting the data). The ``missing'' conductance structure observed experimentally, or rather the ``additional'' fine structure predicted by theory, suggests that certain key ingredients that control the visibility of various conductance features are not properly accounted for by theory. In other words, the experimental resolution in the SM-SC nanowire tunnel conductance measurements is generically much lower than the predictions of the minimal theory. This is in sharp contrast with most solid state  transport measurements where experimental data generically manifest more structures than the theoretical predictions since the theoretical models typically leave out many unknown details of the realistic system. In this section, we explore different possible operational mechanisms which could be responsible for this discrepancy. Specifically, we investigate the role of the effective length of the nanowire, self-energy effect stemming from the proximity to the parent superconductor, finite temperature, finite dissipation, and multi-band effect. Note that no single effect from those mentioned above can fully explain the experimental observations. Instead, a combination of these effects (and perhaps other as yet unknown) is more likely to provide the solution to the apparent discrepancy between experiment and theory. Our investigation is based on an analysis of the main features that characterize the dependence of the low-energy conductance spectra maps on the strength of the applied Zeeman field for different field orientations. This analysis enables us to identify the relationship between the ``visible'' spectral features and certain basic properties of the hybrid system.

\begin{figure}[tbp]
\centering
\includegraphics[width=1.0\columnwidth]{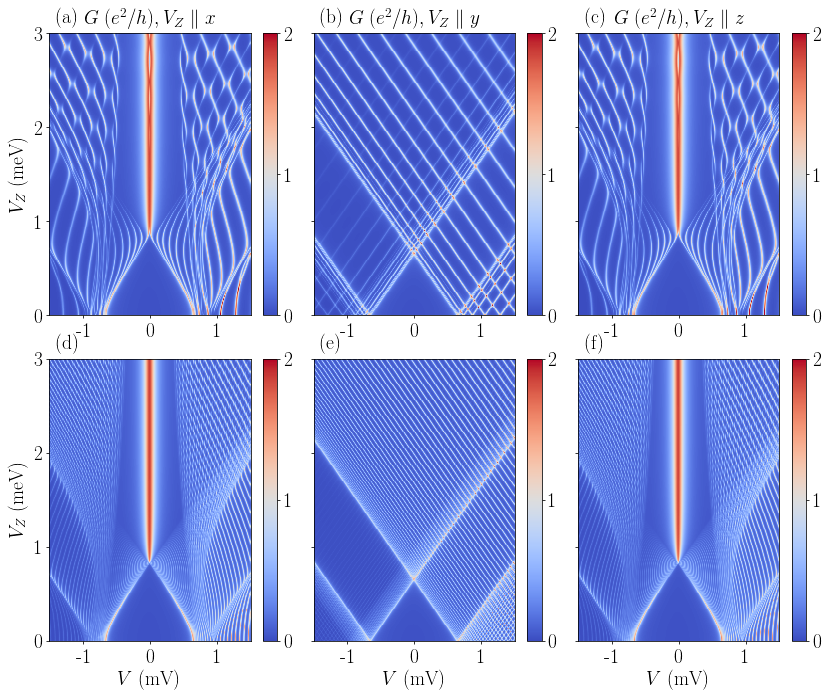}
\caption{Conductance maps for nanowires with short or long effective lengths and different Zeeman field orientations calculated within the weak-coupling theory. The upper panels [(a), (b), (c)] correspond to a short nanowire with $L_{\rm wire}=1.5\mu$m, while the lower panels [(d), (e), (f)] represent  conductance maps for a long nanowire ($L_{\rm wire}=6\mu$m). The left, middle, and right columns correspond to Zeeman fields pointing along the x-, y-, and z-axes, respectively.}
\label{fig:G_length}
\end{figure}

\subsection{Role of effective length}
\label{sec:length}

The nominal length of the SM-SC nanowire can be readily measured although the precise active length operational in the TSC properties may not be easy to discern. By contrast, the SC coherence length of the hybrid nanowire is typically unknown, as the actual values of the effective SC parameters characterizing the nanowire are elusive. For example, the chemical potential of the nanowire in the SM-SC hybrid structure is unknown (and hard to determine). The effective mass of the bare SM wire is also unknown, since it is strongly band-dependent, hence, dependent on the chemical potential. Moreover,  as a result of the proximity effect due to the coupling to the parent superconductor, most effective parameters are strongly renormalized, e.g., the effective mass, the strength of the spin-orbit coupling, and the  Land\'e $g$ factor~\cite{deMoor2018Electric}. Recently, self-consistent  Schr\"odinger-Poisson calculations of SM-SC hybrid structures have shown explicitly  that the renormalization of these parameters can be significant and is sensitively dependent on the back gate voltage and the geometry of the hybrid structure~\cite{Woods2018Effective, Antipov2018Effects, Mikkelsen2018Hybridization}. Consequently, the effective dimensionless length of the wire defined, for example, as $L_{\text{wire}}/\xi$, where $\xi$ is the {\em actual} magnetic field-dependent SC coherence length, could be very different from the estimate based on nominal ``bare'' values of the system parameters. In turn, this affects the density of states of the nanowire, i.e., the energy separation between quantum states that could produce distinct spectral features in a tunneling measurement. This discussion on the serious difficulties in ascertaining even the (dimensionless) length of the nanowire, a seemingly obvious system parameter, demonstrates the problem one faces in modeling the realistic SM-SC structures in the context of Majorana experiments.

Since the values of the effective parameters are difficult to determine (either experimentally or theoretically), we consider  $L_{\text{wire}}$ (rather than $\xi$) in Eq.~\eqref{eq:1d_smsc} as our control parameter and investigate its role in suppressing the emergence of well-defined spectral features. More specifically, in this subsection we calculate the tunnel conductance for nanowires with various effective lengths. In practice,  long (short) nanowires could correspond to systems of a given nominal length and a short (long) SC coherence length. We emphasize that the physics is very different for `long' (i.e., $L_{\text{wire}} \gg \xi$) versus `short' ($L_{\text{wire}} \lesssim \xi$) wires. In the numerical simulation we use $h_{\text{sc}} = \Delta_{\rm ind} \tau_x, \Delta_{\rm ind} = 0.65$meV, and $\mu=0.5$meV. 

Calculated conductance maps showing the dependence of the differential conductance on the strength of the applied Zeeman field and the bias voltage for different wire lengths and field orientations are presented (without any self-energy effect in the theory) in Fig.~\ref{fig:G_length}.
The upper panels are conductance maps for a hybrid nanowire of (nominal) length $L_{\rm wire}=1.5~\mu$m. In Fig.~\ref{fig:G_length}(a), the Zeeman field is oriented along the x-axis, i.e., it is parallel to the wire axis. When the strength of the Zeeman field is larger than the critical field $V_Z > V_{Zc} = \sqrt{\Delta^2 + \mu^2} \simeq 0.82$meV, a ZBCP of height $2e^2/h$ forms signaling the emergence of a MZM at the (left) end of the wire. In the vicinity of $V_{Zc}$, we clearly notice conductance features associated with bulk gap closing and reopening, which indicate the (finite-size remnant of the) TQPT. Thus, the emergence of the quantized ZBCP (associated with the MZM in the topological regime) and the bulk gap closing and reopening (associated with the TQPT), which happen simultaneously, as generically expected, provide a clear signature of the MZM nature of the ZBCP.  Note also that for the short nanowire in Fig.~\ref{fig:G_length}(a) the characteristic finite-size effects also show up. For example, the conductance features  associated with the bulk gap `closing and reopening' are discrete, with the level spacing between bulk states being inversely proportional to the length of the nanowire. This is the direct result of the `size quantization' of the energy levels in the nanowire due to its finite size (i.e., short length) along the x-direction. In addition, there are conductance features above the induced SC gap ($\Delta_{\rm ind} = 0.65~$meV), also showing discretization due to the finite-size effect. Furthermore, in the large Zeeman field regime, i.e., $V_Z > 2~$meV, one can clearly see energy splitting oscillation of the ZBCP characteristic of the wave function overlap between the two MZMs localized at the opposite ends of the nanowire. Both of these subtle conductance structures, the discrete level structures due to size quantization and the Majorana oscillations due to MZM wave function overlap from the two ends, are generic theoretical features in short wires.

When the Zeeman field points along the y-axis, there is no ZBCP and the bulk gap clearly vanishes above a certain critical value of the applied Zeeman field, as shown in Fig.~\ref{fig:G_length}(b). More specifically,  the SC gap closes when $V_Z > \Delta_{\rm ind}$ and never reopens, since the spin-singlet $s$-wave pairing is suppressed by the Zeeman field. All the conductance features in Fig.~\ref{fig:G_length}(b) are associated with bulk SC states and, therefore, show characteristic discrete features due to the finite-size effect. Finally, in Fig.~\ref{fig:G_length}(c), the Zeeman field is oriented along the z-axis and the corresponding conductance map is identical to that in Fig.~\ref{fig:G_length}(a). This equivalence between the x- and the z-orientations is due to the fact that the Hamiltonians corresponding to the two orientations are connected by a unitary rotation $U = e^{i\pi \sigma_y /4}$ in the spin space. In other words, as long as the Zeeman field is normal to the spin-orbit field ($\propto \sigma_y$), the conductance map should remain the same. Note that this symmetry property holds only if we neglect orbital effects (i.e., in the one-dimensional limit). The orbital magnetic field effects are expected to produce some differences between the conductance maps corresponding to the x- and z-orientations, although qualitatively they should remain quite similar, particularly since orbital field effects should be small in a one-dimensional system.

The lower panels of Fig.~\ref{fig:G_length} show the conductance maps for a long nanowire of length $L_{\rm wire}=6\mu$m, i.e., having an effective length four times larger than the wire that generates the maps shown in the upper panels. The dominant feature -- the quantized ZBCP emerging above the critical field associated with the TQPT -- is very similar to that shown in the upper panels, indicating that it is generated by a localized state, which is not affected by changes in the length of the system as long as $L_{\rm wire} \gg \xi$. Note, however, that there is no visible Majorana oscillation induced splitting of the ZBCP at large Zeeman field since the MZM wavefunction overlap from the two ends is suppressed because of the larger separation between the two wire ends. The main effect of increasing the length of the wire is a substantial reduction of the finite-size level spacing between the bulk states doubled by a reduction of the weight of each individual feature associated with a bulk state. This reduction is a consequence of the wave function amplitude of bulk states decreasing with $L_{\rm wire}$, which reduces the effective coupling between these states and the tunnel probe. As a consequence, the features associated with bulk states tend to become a smoothly varying background with significantly less structure than its short-wire counterpart. The remaining well-defined features, e.g., the feature associated with the closing of the quasiparticle gap at the TQPT, are generated by states localized at the left end of the wire, rather that bulk states. The gap closing feature, for example, is produced by an intrinsic ABS, which emerges generically in pristine wires with finite chemical potential~\cite{Huang2018Metamorphosis}. Note that there is no equivalent well-defined feature associated with the reopening of the quasiparticle gap. 

Our conclusion is that the conductance maps measured in the short-wire and long-wire regimes have qualitatively different properties regarding the visibility of bulk states (i.e., states that extend throughout the whole system): while in short wires the bulk states can produce well-defined, separated features, in the long-wire regime  they generate a smoothly varying, structureless background. In long wires, all well-defined features  are associated with states localized near the end of the system, which couple strongly with the tunnel probe. There is a characteristic effective length that separates the short-wire and long-wire regimes. To identify this length scale, which depends crucially on the induced superconducting coherence length in the nanowire, it is essential to know the effective system parameters, particularly the value of the effective mass. Thus, in principle, the absence of conductance structures in the experiment could arise from the experimental nanowire samples somehow being in the long-wire regime of $L_{\rm wire} \gg \xi$ (in spite of their physical lengths being relatively short $\sim$ 1$\mu$m).

\begin{figure}[tbp]
\centering
\includegraphics[width=1.0\columnwidth]{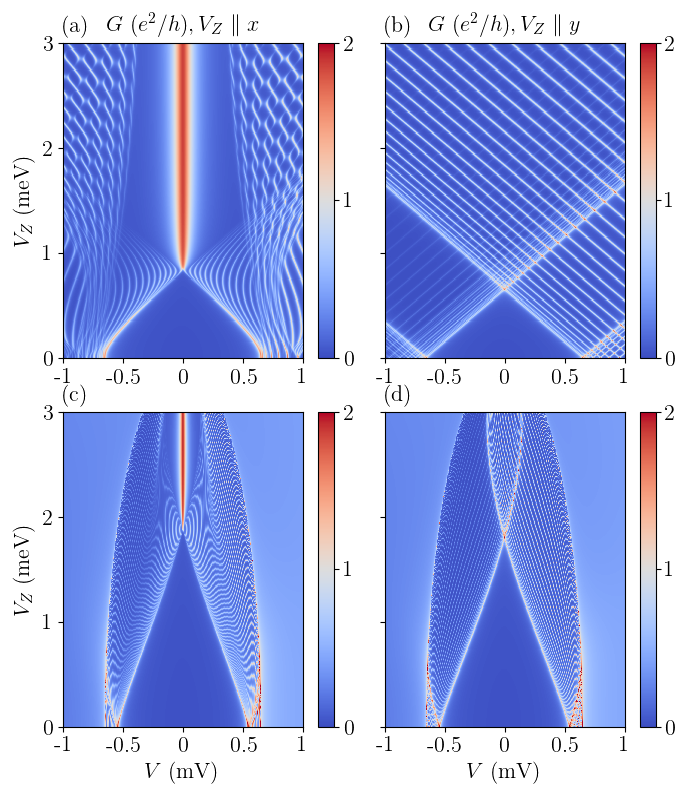}
\caption{Conductance for nanowires of $L_{\text{wire}}=3\mu$m, $\mu=0.5$meV with different models of proximity superconductivity. In (a) and (b), the SM-SC interface is in the weak-coupling limit. The proximity effect is modeled by an induced SC pairing term $h_{\rm sc} = \Delta_{\rm ind} \tau_x$, with $\Delta_{\rm ind} = 0.65$meV. In (c) and (d), the SM-SC interface is in the strong-coupling limit. The proximity effect is modeled as a self-energy term $h_{\rm{sc}} = \Sigma(\omega)$, with effective coupling $\lambda=1.8$meV. In (a) and (c), the Zeeman field points along the x-axis, while in (b) and (d) the Zeeman field points along the y-axis.  }
\label{fig:G_se}
\end{figure}

\subsection{Self-energy effect}

The hybrid SM-SC system is composed of a spin-orbit coupled semiconducting nanowire in proximity with a conventional $s$-wave superconductor. In the previous section, we have treated the proximity effect induced by the parent SC within the weak-coupling approximation, using the effective pairing term $h_{\text{sc}} = \Delta_{\rm ind} \tau_x$. 
This approximation is appropriate for describing low-energy (i.e., low-bias voltage) features when the effective SM-SC coupling is much smaller than the gap of the parent SC, i.e., $\lambda \ll \Delta$. However, if the SM-SC coupling is strong and/or we are concerned about energies comparable to the induced gap, the parent SC cannot be treated as a simple source of Cooper pairs, but has to be modeled in more detail. In realistic experiments, the effective SM-SC coupling is an unknown parameter which sensitively depends on the profile of the wave function through some external conditions, e.g., the back gate voltage~\cite{deMoor2018Electric}. Specifically, one can integrate out the degrees of freedom of the parent SC and incorporate its effect on the SM wire as an interface self-energy contribution~\cite{Stanescu2010Proximity, Stanescu2017Proximity}. Using a simple mean-field model for the parent SC generates a self-energy of the form
\begin{align}
&h_{\text{sc}} = \Sigma(\omega) = -\lambda \frac{\omega \tau_0 + \Delta \tau_x}{\sqrt{\Delta^2-\omega^2}}, \nn
& \Delta = \Delta_0 \sqrt{1-(V_Z/V^*_{Z})^2},
\end{align}
where $\omega$ denotes the energy ($\hbar=1$), $\lambda$ is the effective SM-SC coupling producing the proximity effect, and $\Delta_0$ is the bulk SC gap of the parent superconductor at zero magnetic field. We assume that the gap of the parent superconductor shrinks with increasing strength of the Zeeman field and is completely suppressed above $V_Z = V^*_Z$. This reduction of the parent SC gap with increasing field is always observed in nanowire experiments, where the SC gap collapses at some finite magnetic field value of 1 - 2 T.

The conductance maps for a system with $L_{\text{wire}}=3~\mu$m,  $\mu=0.5~$meV, $\Delta_0=0.65~$meV, $\lambda=1.8~$meV, and $V^*_Z=3.5~$meV are shown in Fig.~\ref{fig:G_se}(c) and (d). For comparison, in the upper panels, Fig.~\ref{fig:G_se}(a) and (b), we show the corresponding maps for a similar weakly coupled system with $\Delta_{\rm ind}=0.65~$meV. A notable characteristic of the spectra obtained within the self-energy approach is the presence of a continuum above the (field-dependent) parent SC gap. This feature is generated by the continuum of quasiparticle states in the parent SC. At energies larger than the parent SC gap $\Delta$, there is no discrete state in the SM nanowire, since all the SM states are hybridized with the parent SC (quasi)continuum. Consequently, the conductance map becomes structureless above the parent SC gap. Note that the conductance calculations shown in Fig.~\ref{fig:G_se}  include only contributions from Andreev processes. However, when $|V|>\Delta$, there are also contributions from finite energy quasiparticles  excited inside the parent SC. Including these contributions enhances the differential conductance at bias voltages above the parent SC gap and reveals characteristic features, such as the SC coherence peak~\cite{Reeg2017Transport, Stenger2017Transport}.

The most significant characteristic of the conductance maps obtained within the self-energy approach stems from the proximity-induced low-energy renormalization that affects the subgap features, i.e., those corresponding to $|V| < \Delta$. First, we note that the level spacing associated with the finite length of the wire is significantly reduced (as compared level spacing in the upper panels). This can be naturally understood as a proximity-induced renormalization of the effective mass. Most importantly, the critical fields $V_{Zc}^x$ and $V_{Zc}^y$ associated with the closing of the bulk gap in systems with Zeeman field oriented along the x- and y-directions, respectively, are not controlled by the induced gap ($\Delta_{\rm ind})$, but by the strength of the effective SM-SC coupling ($\lambda$), as evident in Fig.~\ref{fig:G_se}(c) and (d). This observation leads us to the key question: what information can one extract from  measurements of the gap-closing features along two different directions? If we assume a pristine (i.e., uniform) system within the independent-band approximation (which holds when the inter-band spacing is much larger than other relevant energy scales), the key parameters that control the closing of the bulk gap are the effective SM-SC coupling, $\lambda$, and the  chemical potential, $\mu$. In addition, one has to keep in mind that the experimentally accessible field parameter is the magnetic field $B$, rather than the Zeeman splitting $V_Z$, hence, the effective $g$ factor is the third key parameter. On the other hand, orientation-dependent gap closing measurements give us access to three relatively robust and well-defined quantities: the induced SC gap $\Delta_{\rm ind}$ (at zero magnetic field and inside the quasiparticle continuum) and the critical fields $B_c^x$ and $B_c^y$ associated with the closing of the bulk gap for the corresponding field orientations. The relations between the experimentally-accessible quantities and the relevant system parameters are~\cite{Stanescu2017Proximity}
\begin{eqnarray}
\Delta_{\rm ind}\sqrt{\Delta_0 +\Delta_{\rm ind}} &=& \lambda \sqrt{\Delta_0 -\Delta_{\rm ind}}, \label{parameters1} \\
g_{\rm eff}\mu_B B_c^y &=& \lambda, \label{parameters2}\\
g_{\rm eff}\mu_B B_c^x &=& \sqrt{\mu^2 + \lambda^2},  \label{parameters3}
\end{eqnarray}
where $\mu_B$ is the Bohr magneton. The first equation provides an estimate of the effective SM-SC coupling, the second equation gives the effective $g$ factor, while the third equation allows one to extract the chemical potential in the nanowire. Note that in the weak-coupling limit ($\lambda \ll \Delta_0$) we have $\Delta_{\rm ind}\approx \lambda$. In general, the effective $g$ factor can be extracted  from a measurement of the gap suppression in the presence of a field oriented along the y-direction [see, for example, Fig.~\ref{fig:G_se}(b)]. However, the minimal information that allows the extraction of all three parameters, $g_{\rm eff}$, $\lambda$ and $\mu$, can only be obtained by combining the conductance maps corresponding to the $x$ and $y$ field orientations. We emphasize that Eqs.~(\ref{parameters1}-\ref{parameters3}) provide a first approximation for the effective parameters, which is valid in pristine systems if orbital and band-mixing effects are negligible. Under these assumptions, using the experimental data in Fig.1 of Ref.~\cite{Bommer2018SpinOrbit} (e.g., $\Delta_0 = 0.65~$meV, $\Delta_{\rm ind}=0.49~$meV, $B^y_c=0.25~$T, $B^x_c=0.7~$T), we estimate that $\lambda \approx 1.34~$meV, $g_{\rm eff} \approx 92.6$, and $\mu \approx 3.5~$meV. Including other effects such as the orbital and band-mixing effects requires more microscopic approaches that treat the semiconductor nanowire and the parent superconductor on equal footing using  two- or three-dimensional tight-binding models~\cite{Woods2018Effective, Antipov2018Effects, Mikkelsen2018Hybridization}. Such numerical calculations are beyond the scope of this work, and also necessitate a knowledge of other bulk parameters and various boundary conditions which are not available experimentally.



\subsection{Finite temperature and dissipation}

The level spacing between bulk states is controlled by the effective length of the wire and by the strength of the SM-SC coupling, as discussed in the previous subsections. For a given level spacing, the visibility of bulk features is determined by the characteristic broadening of these features, which depends on the effective coupling between the bulk states and the tunnel probe, and by finite temperature and the possible presence of dissipation. A broadening comparable to (or larger than) the level spacing results in the smoothening of the (bulk) conductance features and the suppression of well-defined (bulk-induced) conductance structures. Essentially, any physical mechanism causing level broadening will suppress and smoothen fine structures in the measured conductance. 

\begin{figure}[tbp]
\centering
\includegraphics[width=1.0\columnwidth]{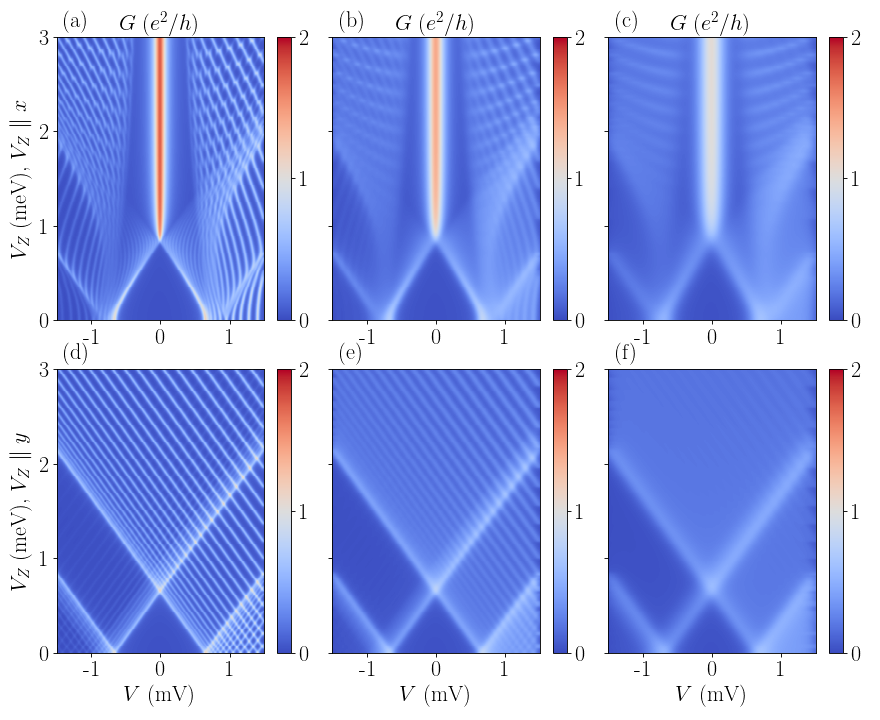}
\caption{Conductance maps for Majorana nanowires at different (effective) electron temperatures. The upper panels correspond to a Zeeman field pointing along the x-axis, while the lower panels are for a Zeeman field pointing along the y-axis. In the left, middle, and right columns, the temperatures are $T=0.01, 0.03$, and $0.06~$meV, respectively.}
\label{fig:G_T}
\end{figure}

In a tunneling experiment, although the nominal fridge temperature is known, the actual temperature of the electrons inside the sample, which may be much higher, is typically unknown. This is a rather well-known problem in semiconductors where electrons often do not come to equilibrium with the lattice at low lattice temperatures. Here, we consider the effect of finite temperature on the conductance spectral structure under the assumption that the effective temperature is much higher that the nominal temperature of the fridge. The tunnel conductance at finite temperature is calculated as the convolution between the zero-temperature conductance $G_0(V)$ and the derivative of Fermi function $f^\prime_{\text{T}}(E)$:
\begin{align}
G_{\text{T}}(V) = -\int^{+\infty}_{-\infty} dE G_0(V-E) f^\prime_{\text{T}}(E).
\end{align}
We consider a nanowire with effective parameters  $L_{\rm wire}=3~\mu$m, $\mu=0.5~$meV, and $\Delta_{\rm ind}=0.65~$meV and calculate the conductance maps for three different temperatures: $k_BT = 0.01, 0.03, 0.06~$meV, where $k_B$ is the Boltzmann constant. The results are shown in Fig.~\ref{fig:G_T}. As the temperature increases, the conductance becomes increasingly structureless and the features associated with bulk states are smeared out. By contrast, the features associated with states localized at the end of the wire  remain clearly visible, although they get broadened and their height is reduced. The main features associated with localized states survive including the MZM-induced ZBCP and the gap closing feature generated by intrinsic ABSs.


\begin{figure}[tbp]
\centering
\includegraphics[width=1.0\columnwidth]{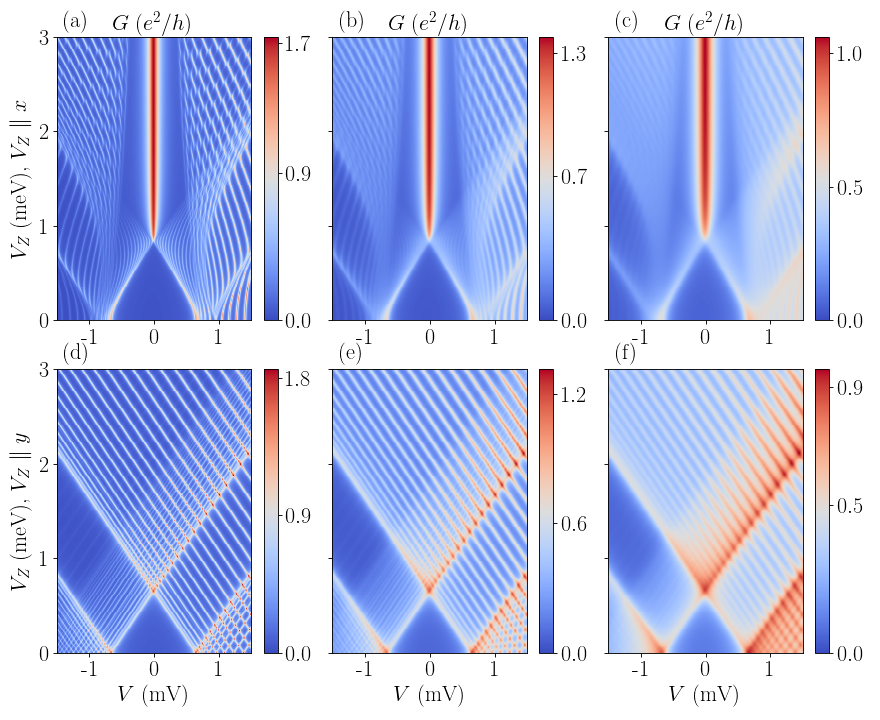}
\caption{Conductance maps for Majorana nanowires in the presence of different amounts of dissipation. The upper panels correspond to a Zeeman field pointing along the x-axis, while the lower panels are for a Zeeman field pointing along the y-axis. In the left, middle, and right columns, the dissipation strengths are $\Gamma=0.01, 0.03$, and $0.06~$meV, respectively.}
\label{fig:G_dissipation}
\end{figure}

Similar to finite temperature, dissipation can also smoothen the fine structures in the conductance maps. In real systems, dissipation may arise from disorder-induced low-energy states in both the SM nanowire and the parent SC, as well as vortices in the parent superconductor~\cite{Liu2017Role}. In addition, dissipation may result from the so-called inverse proximity effect induced by the coupling to the normal-metal lead~\cite{Danon2017Conductance}. In this work, we do not consider a specific source of dissipation, but rather model it phenomenologically by adding an imaginary (dissipation) term $i\Gamma$ in the BdG Hamiltonian, as shown in Eq.~\eqref{eq:1d_smsc}. The nanowire parameters used in the calculation are: $L_{\rm wire}=3~\mu$m, $\mu=0.5~$meV, $\Delta_{\rm ind}=0.65~$meV. We calculate the conductance maps for three different dissipation strengths: $\Gamma = 0.01, 0.03, 0.06~$meV. The results are shown in Fig.~\ref{fig:G_dissipation}. 
The basic effect of dissipation is quite similar to that of finite temperature: it broadens the spectral features and, as a result, suppresses the fine conductance structures. In addition, there is a feature that is uniquely related to the presence of dissipation: the breaking of the particle-hole symmetry and the emergence of ``stripy'' features in conductance maps. When the superconducting nanowire is coupled to dissipative sources (e.g., disorder-induced low energy states, normal-metal lead, and extra fermionic bath.), the tunnel conductance will be proportional to particle (hole) component of the eigenstate at positive (negative) bias voltage~\cite{Martin2014Noneq}. We emphasize that our treatment of dissipation is in a phenomenological way, i.e., by adding an imaginary term in the BdG Hamiltonian of the Majorana nanowire. Any microscopic analysis of the dissipation sources is beyond the scope of the current work. We note that the apparent particle-hole asymmetry in Fig.~\ref{fig:G_dissipation} indicates quasiparticle excitations having a predominantly particle (or hole) character. By contrast, BdG states characterized by an approximately equal mix of particle and hole components generate particle-hole symmetric features even in the presence of dissipation~\cite{Liu2017Role}. 



\subsection{Multiband effect}

So far, we have only considered single-band models. However, in general one would expect the hybrid structures to be characterized by multiband occupancy. In this subsection we investigate specific signatures characterizing the low-energy differential conductance maps that stem from the multiband character of the hybrid system. We note that the relevant bands that play a role in the low-energy physics of SM-SC nanowires correspond to different confinement-induced conduction subbands. In principle, the effective model parameters, such as effective mass, spin-orbit coupling, Land\'e $g$ factor, and induced SC pairing gap, are band-dependent. However, incorporating this elaborate band dependence of the microscopic parameters is beyond the scope of this work. In addition,  the occupancy of different  transverse modes depends on the gate voltage in a nonlinear manner~\cite{Woods2018Effective, Antipov2018Effects, Mikkelsen2018Hybridization} and is unknown experimentally in the SM-SC nanowires. Here, we consider a single spinful channel in the normal-metal lead coupled to several partially occupied  subbands in the nanowire, as shown in  Fig.~\ref{fig:schematic}(c). For simplicity, the only difference between the subbands that we explicitly account for is the chemical potential (measured relative to the spin-orbit crossing point of each band); all other parameters are assumed to have band-independent values. The values of the chemical potential used in the numerical calculations are $\mu=1,3$, and $5~$meV for the lowest three subbands, respectively. This simple physical model is adequate for the purpose of the qualitative physics we are interested in.

\begin{figure}[tbp]
\centering
\includegraphics[width=1.0\columnwidth]{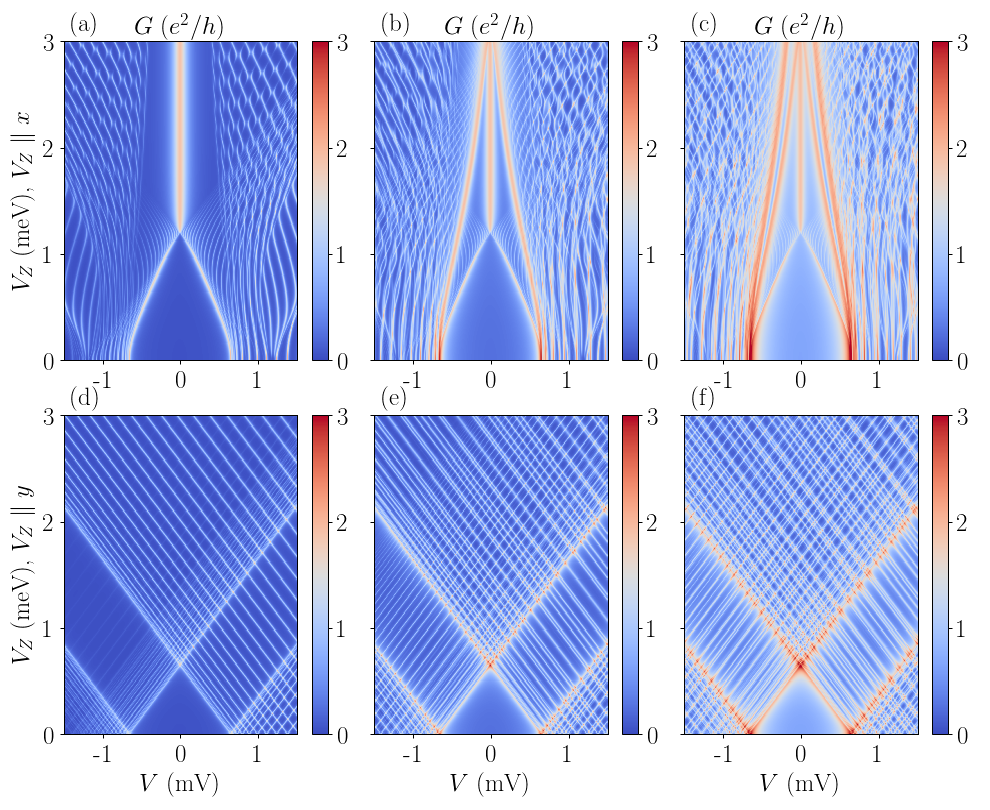}
\caption{Differential conductance as a function of the strength of the Zeeman field for Majorana nanowires with different number of subbands. The upper panels correspond to a Zeeman field pointing along the x-axis, while the lower panels are for a Zeeman field pointing along the y-axis. The left, middle, and right columns correspond to systems with one, two, and three subbands, respectively. The values of the chemical potential (measured relative to the spin-orbit crossing point of each band) are $\mu=1,3,$ and $5~$meV for the lowest three subbands. }
\label{fig:G_subband}
\end{figure}

Differential conductance maps for nanowires with different numbers of subbands are shown in Fig.~\ref{fig:G_subband}.  The upper panels correspond to a  Zeeman field pointing along the x-axis, i.e., parallel to the wire axis, while the lower panels  are for Zeeman fields pointing along the y-axis, i.e., parallel to the spin-orbit field. We consider systems with two [Figs.~\ref{fig:G_subband}(b) and (e)] and three  [Figs.~\ref{fig:G_subband}(c) and (f)] partially occupied subbands; for comparison, we also include the single-band results [Figs.~\ref{fig:G_subband}(a) and (d)].

The conductance maps for a two-band system are shown in  Figs.~\ref{fig:G_subband}(b) and (e). As the strength of the Zeeman field increases [see Fig.~\ref{fig:G_subband}(b)], the bulk gap closes at $V_{Zc}\simeq 1.2~$meV, as signaled by the gap-closing feature generated by the intrinsic ABS associated with the top (Majorana) band. For $V_Z>V_{Zc}$ a Majorana-induced ZBCP is clearly visible. These features, which are generated by states associated with the top band, are similar to the corresponding feature in a single-band system [see Fig.~\ref{fig:G_subband}(a)]. However, in contrast with the single-band scenario, in Fig.~\ref{fig:G_subband}(b) one can clearly distinguish a second strong feature that converges slowly toward zero energy at $V_Z>3~$meV. This second feature is generated by the (intrinsic) ABS associated with the low-energy band. In addition, the second band also introduces bulk states, reducing the level spacing (as compared with the single-band case). On the other hand, when the field is oriented along the y-direction [see Fig.~\ref{fig:G_subband}(e)], there is no qualitative difference between multi-band and single-band scenarios [Figs.~\ref{fig:G_subband}(d) and (e)]. Note that this type of  behavior implies that the system parameters (e.g., the induced gap and the effective $g$ factor) are band-independent (or, in practice, weakly band-dependent). The multiband features discussed above, i.e., the additional conductance peaks generated by intrinsic ABSs~\cite{Huang2018Metamorphosis} associated with occupied low-energy bands and the reduced level spacing between bulk states, are also present in the conductance map of a system with three occupied bands, as shown in Figs.~\ref{fig:G_subband}(c) and (f). As a general trend, we notice that, as the number of occupied bands increases, the relative weight of in-gap features associated with the top occupied band (e.g., the Majorana ZBCP) decreases with respect to the weight of the weakly field-dependent features associated with low-energy bands.

\begin{figure}[tbp]
\centering
\includegraphics[width=1.0\columnwidth]{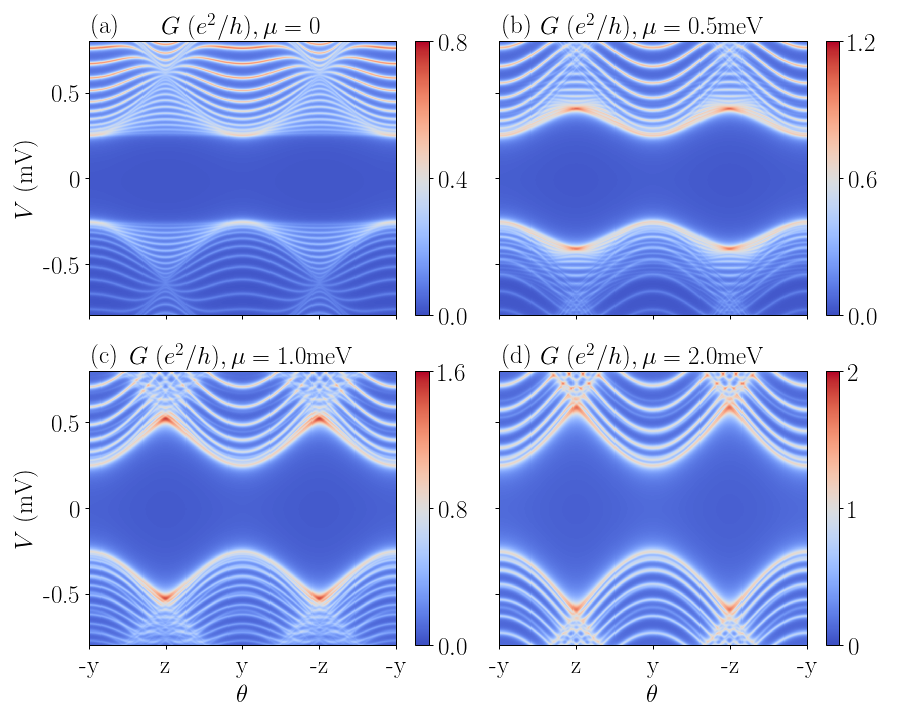}
\caption{Conductance map as a function of the magnetic field orientation (see Fig.~\ref{fig:schematic} for the definition of the angle $\theta$) for a pristine hybrid system with parameters $V_Z=0.4~$meV, $\Delta_{\rm ind}=0.65~$meV, $L_{\text{wire}} = 3~\mu$m, $\alpha_R = 0.5~$eV$\angstrom$ and different values of the chemical potential: (a) $\mu= 0$, (b) $\mu=0.5~$meV, (c) $\mu=1~$meV, and (d) $\mu=2~$meV.
The strength of the Zeeman field is fixed to be less than the critical value $V_{Zc}^y$ corresponding to the vanishing of the bulk gap. Note that, in the weak-coupling limit, $V_{Zc}^y=\Delta_{\rm ind}$.}
\label{fig:RotationSmallVz}
\end{figure}


\subsection{Continuous rotation of the magnetic field} \label{sec:rotation}

In this subsection, we discuss the conductance maps of pristine semiconductor-superconductor nanowires obtained by continuously rotating the Zeeman field. This discussion, included here for completeness, concerns an aspect of critical importance in experiment: the actual orientation of the effective spin-orbit field. Of course, in model calculations the spin-orbit coupling is included ``by hand'' and the orientation of the corresponding effective field is known (e.g., along the y-axis in our model). However, in a more microscopic approach that includes the self-consistent calculation of electrostatic effects -- which, in turn, will determine the spin-orbit coupling -- and, especially, in experiment, performing a full rotation of the magnetic field to identify the direction of the spin-orbit field is essential. Experimentally, the precise direction of the spin-orbit field is unknown whereas in the minimal model it is stipulated to be along the y-direction.

As discussed in Sec.~\ref{sec:length}, Zeeman fields pointing along an arbitrary direction in the x-z plane are equivalent within our model. Therefore we restrict our analysis to rotations within the y-z plane (i.e., the plane perpendicular to the wire), as done in Ref.~\cite{Bommer2018SpinOrbit}. By definition, $\theta$ is the angle between the Zeeman field and z-axis, as shown in  Fig.~\ref{fig:schematic}. The model parameters used in the calculation are $L_{\rm wire}=3~\mu$m and $\alpha_R$ = 0.5~eV$\angstrom$.

First, we consider the weak Zeeman field regime corresponding to $V_Z<V_{Zc}^y$, where $V_{Zc}^y = \Delta_{\rm ind}$ is the critical field associated with the closing of the bulk gap for a system with a Zeeman field oriented parallel to the effective spin-orbit field. The system parameters are $V_Z=0.4~$meV, $\Delta_{\rm ind}=0.65~$meV, $L_{\text{wire}} = 3~\mu$m, $\alpha_R = 0.5~$eV$\angstrom$ and four different values of the chemical potential,  $\mu=0, 0.5, 1$, and $2~$meV. The conductance as a function of the angle $\theta$ is shown in Fig.~\ref{fig:RotationSmallVz}. A common feature in the four panels (which correspond to different values of the chemical potential) is the finite SC gap that persists at all angles. The minimum gap corresponds to $\theta =\pm\pi/2$, i.e., for a Zeeman field parallel to the spin-orbit field (pointing along the y-axis). In the weak-coupling limit, the magnitude of the minimum gap is determined by $\Delta_{\rm ind} - V_Z$. On the other hand, the maximum gap, which obtains along the z-direction (i.e., perpendicular to the spin-orbit field), depends on the chemical potential, increasing with $\mu$. As a result, the gap, which is  isotropic at $\mu=0$, becomes more anisotropic with increasing chemical potential, as shown in Fig.~\ref{fig:RotationSmallVz}.

\begin{figure}[tbp]
\centering
\includegraphics[width=1.0\columnwidth]{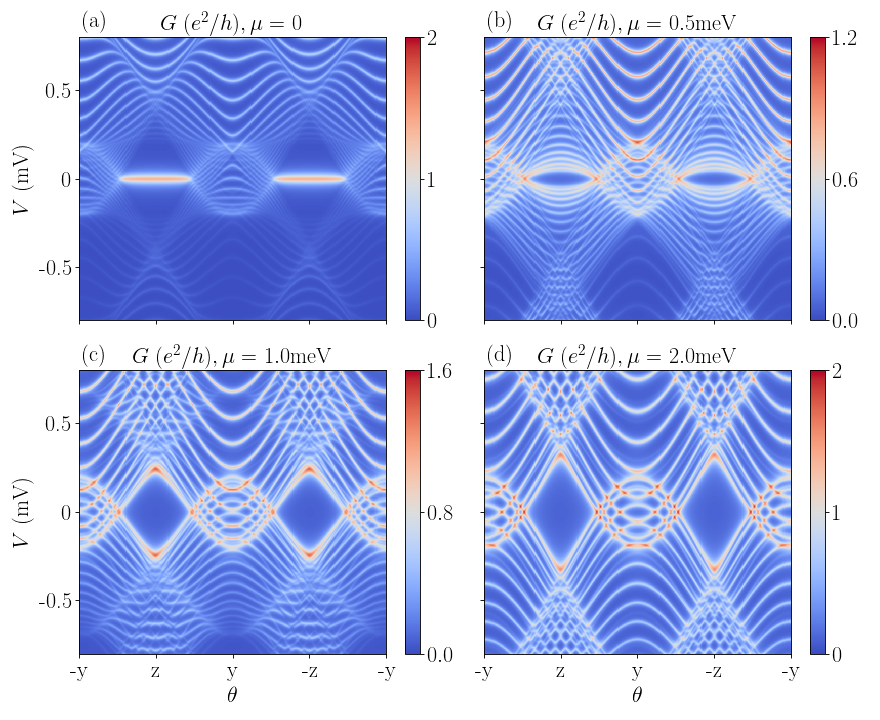}
\caption{Conductance map as a function of the magnetic field orientation (see Fig. \ref{fig:schematic} for the definition of the angle $\theta$) for a pristine hybrid system with parameters $V_Z=0.6~$meV, $\Delta_{\rm ind}=0.4~$meV, $L_{\text{wire}} = 3~\mu$m, $\alpha_R = 0.5~$eV$\angstrom$ and different values of the chemical potential: (a) $\mu= 0$, (b) $\mu=0.5~$meV, (c) $\mu=1~$meV, and (d) $\mu=2~$meV.
Note that the strength of the Zeeman field is fixed to a values larger than $V_{Zc}^y = \Delta_{\rm ind}$.}
\label{fig:RotationLargeVz}
\end{figure}

Next, we consider the strong Zeeman field regime corresponding to $V_Z>V_{Zc}^y$. The system parameters are $V_Z=0.6~$meV, $\Delta_{\rm ind}=0.4~$meV, $L_{\text{wire}} = 3~\mu$m, $\alpha_R$ = 0.5~eV$\angstrom$ and we consider four different values of chemical potential, $\mu=0, 0.5, 1$, and $2~$meV. The conductance as a function of the angle $\theta$ is shown in Fig.~\ref{fig:RotationLargeVz}. Note that in the vicinity of $\theta=\pm\pi/2$ (y-axis), the spectrum is gapless. On the other hand, in the vicinity of $\theta=0/\pi$ (z-axis), there is a finite gap, possibly with a midgap ZBCP, if the topological condition $V_Z > V_{Zc}^x$ is satisfied. Indeed, when the chemical potential is small [e.g., $\mu=0$ in Fig.~\ref{fig:RotationLargeVz}(a)], a ZBCP forms inside the gap when the Zeeman field is approximately oriented along the z-axis. This ZBCP is MZM-induced  in the topological phase with $V_Z > V_{Zc}^x$. In Figs.~\ref{fig:RotationLargeVz}(b-d), on the other hand, there is no ZBCP because the system is still in the topologically trivial regime. 




\section{Anisotropic suppression of induced superconductivity in non-homogeneous hybrid systems} \label{sec:ABS}

\begin{figure}[tbp]
\centering
\includegraphics[width=1.0\columnwidth]{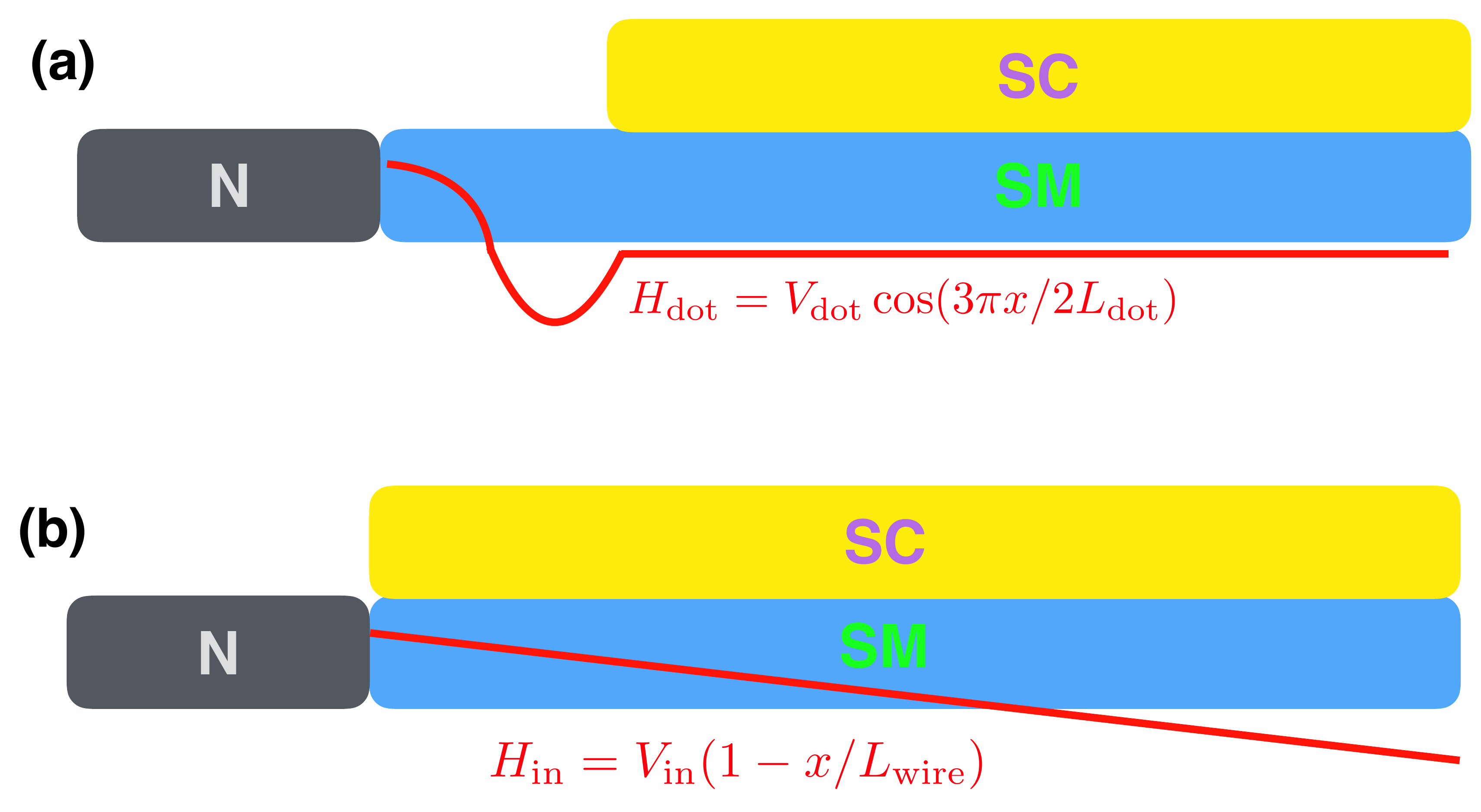}
\caption{Schematic representation of nanowires in the presence of (a) an external quantum dot or (b) a smooth chemical potential inhomogeneity. (a) The (proximitized) nanowire is coupled to a quantum dot defined by the uncovered region  between the normal-metal lead and the hybrid nanowire. The effective potential in the dot region has a  profile given by the effective potential $h_{\text{dot}}=V_{\text{dot}} \cos ( 3\pi x / 2L_{\text{dot}} )$ (red line). (b) Nanowire with smoothly varying chemical potential inhomogeneity. The profile of the smooth potential is given by the linear profile $h_{\text{in}}=V_{\text{in}} ( 1 - x/L_{\text{wire}} )$, as shown by the red line.}
\label{fig:Schematic_smooth}
\end{figure}

In the previous sections, we have considered the orientation-dependent suppression of induced SC gap by the applied Zeeman field in clean,  homogeneous hybrid systems. In such pristine nanowires, the only low-energy subgap states are the Majorana zero modes (in the topological regime, $V_Z>V_{Zc}$) and the intrinsic ABSs (in the trivial regime, $V_Z<V_{Zc}$), which generate characteristic gap-closing features~\cite{Huang2018Metamorphosis}. However, it was shown that the presence of  inhomogeneous effective potentials (e.g., smooth confining potentials, quantum dots near the end of the wire, or chemical potential inhomogeneities) can  induce topologically trivial near-zero-energy extrinsic Andreev bound states~\cite{Kells2012Near, Prada2012Transport,Liu2017Andreev,Chiu2017Conductance,Setiawan2017Electron,Moore2018a,Liu2018Distinguishing,Moore2018b,Vuik2018Reproducing}. 
 These trivial ABSs, which appear, typically, in a parameter regime characterized by large values of the chemical potential, can mimic the MZM phenomenology, including the emergence of robust quantized ZBCPs in local charge tunneling measurements~\cite{Liu2017Andreev}. In this section, we investigate the effect of 
 these low-energy topologically trivial states on the conductance features that signal the  suppression of induced SC gap for various orientations of the applied Zeeman field. In particular, we consider the key question that we want to address in this work: can one discriminate between MZM-induced features and features produced by topologically trivial ABSs based on the dependence of gap closing signatures on the orientation of the applied magnetic field? We focus on non-homogeneous systems characterized by the presence of  an external quantum dot or a chemical potential inhomogeneity (see Fig.~\ref{fig:Schematic_smooth}). To illustrate the fundamental difference between topologically protected MZMs and trivial low-energy ABSs, we also investigate the real-space and spin properties of the wave functions characterizing  the near-zero-energy ABSs.

\subsection{Nanowire coupled to a quantum dot}

We first consider the case of a hybrid nanowire coupled to an external quantum dot defined by the smooth potential profile shown in  Fig.~\ref{fig:Schematic_smooth}(a). The quantum dot lies between the normal-metal lead and the semiconductor-superconductor nanowire. In the experimental setup, such an external quantum dot may be part of the tunnel barrier region, which is not covered by the parent SC or the normal metal. The BdG Hamiltonian describing the quantum dot is
\begin{align}
h_{\text{dot}} = V_{\text{dot}} \cos ( 3\pi x / 2L_{\text{dot}} )\tau_z, \quad 0 \leq x \leq L_{\text{dot}},
\end{align}
where $V_{\text{dot}}$ is the dot potential amplitude and $L_{\text{dot}}$ is the length of the quantum dot. In our simulation, we choose $V_{\text{dot}}=5~$meV and $L_{\text{dot}}=0.3~\mu$m.  The proximitized nanowire is described by the Hamiltonian in Eq.~\eqref{eq:1d_smsc} with  $L_{\rm wire}=3~\mu$m, $\alpha_R=0.5~$eV\angstrom, $\Delta_{\rm ind}=1~$meV, and a (homogeneous) chemical potential $\mu=4.5~$meV. The total Hamiltonian is the sum of the quantum dot and nanowire Hamiltonians. The schematic representation of the dot-nanowire system is shown in Fig.~\ref{fig:Schematic_smooth}(a).

\begin{figure}[tbp]
\centering
\includegraphics[width=1.0\columnwidth]{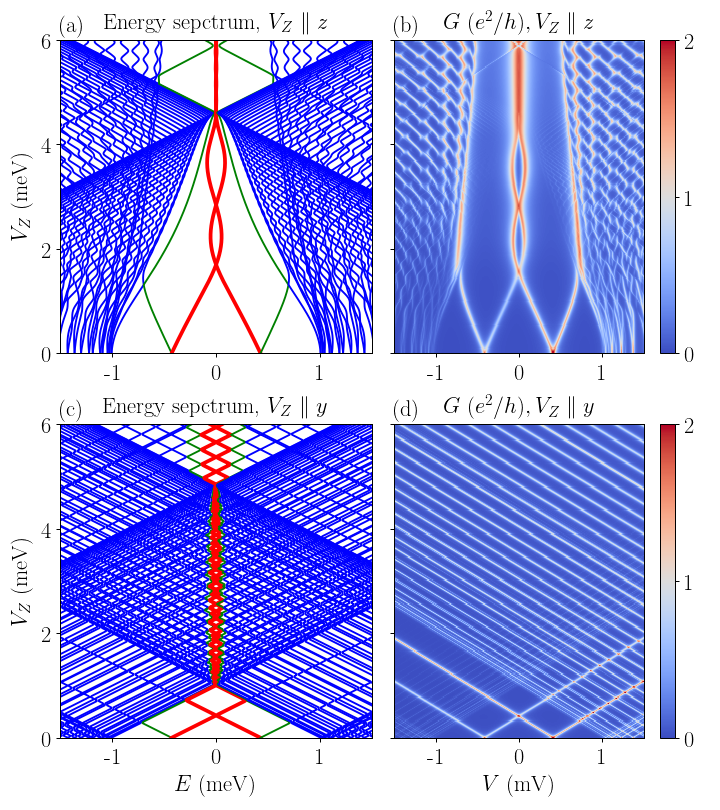}
\caption{Energy spectra (left column) and conductance maps (right column) for the dot-nanowire system. The model parameters used in the calculation are: $L_{\rm wire}=3~\mu$m, $\alpha_R=0.5~$eV\angstrom, $\Delta_{\rm ind}=1~$meV, $\mu=4.5~$meV, $L_{\text{dot}}=0.3~\mu$m, and $V_{\text{dot}}=5~$meV. In the upper (lower) panels, the Zeeman field points along the z-axis (y-axis). The lowest (second-lowest) lying modes are marked in red (green) in the energy spectra (left panels).}
\label{fig:EGdotABS}
\end{figure}

The energy spectra and the conductance maps for the dot-nanowire system are shown in Fig.~\ref{fig:EGdotABS}. Fig.~\ref{fig:EGdotABS}(a) shows the energy spectrum  as a function of the strength of the Zeeman field for a field oriented along the $z$-axis. Note that for $V_Z=0$ there is a Kramer's pair of  ABSs with energy below the induced SC gap. This is a clear signature of the ABS having a reduced pairing potential as a result of 
being localized in the dot region, which  is not proximitized by the parent superconductor. 
Upon increasing the Zeeman field, the ABS localized in the dot region collapses toward zero energy at a Zeeman field way bellow the critical value corresponding to the TQPT  (red line in Fig.~\ref{fig:EGdotABS}), then, for $2 < V_Z < 4~$meV, it oscillates about zero energy with an amplitude significantly smaller than the induced quasiparticle gap. 
In the vicinity of the critical value $V_Z \simeq V_{Zc} \simeq 4.5$meV, one clearly sees a gap closing and reopening feature associated with the vanishing of the bulk gap at the TQPT. For $V_Z > V_{Zc} \simeq 4.5~$meV, a midgap MZM emerges, while around  $V_Z  \simeq 5.8~$meV another ABS localized in the dot region crosses zero energy. The corresponding conductance map is shown in Fig.~\ref{fig:EGdotABS}(b). Note that the signatures  corresponding to the low-energy  ABS as well as the MZM are strong, because both the ABS and the (left) MZM are end states localized in the vicinity of the the normal lead, thus having a large tunnel coupling to the lead. By contrast, the conductance features corresponding to the gap closing and reopening at the TQPT are almost invisible. As  mentioned before, the associated states are bulk states with small characteristic wave vectors ($k\simeq 0$) and, consequently, their wave functions have negligible amplitude at the wire end and do not penetrate significantly inside the dot region. As a result, the effective coupling of these states to the tunnel probe is weak and the corresponding  contributions to the tunneling spectrum are negligible (practically ``invisible''). Note that the signatures visible {\em above} the induced gap are also generated by bulk states. However, these are states with large characteristic wave vectors ($k \simeq k_F$) and, consequently, their wave function amplitudes at the wire end are significantly larger. Of course, proximity-induced re-normalization effects, finite temperature/dissipation, or multi-band effects can suppress or even completely obscure the signatures of these bulk states, as discussed in Sec.~\ref{sec:suppression}.



Next, we consider a dot-nanowire system with a Zeeman field applied along the $y$-axis. The energy spectrum and the corresponding conductance map are shown in Fig.~\ref{fig:EGdotABS}(c) and (d), respectively. The bulk gap vanishes at $V_Z = \Delta_{\rm ind}=1~$meV and never reopens, as shown in Fig.~\ref{fig:EGdotABS}(c). However, the in-gap ABS localized in the dot region crosses zero energy at a Zeeman field $V_Z < \Delta_{\rm ind}$ (red lines). 
Note that the calculation shown in Fig.~\ref{fig:EGdotABS} is valid in the weak-coupling limit. When the SM-SC effective coupling is comparable to (or larger than) the parent SC gap $\Delta_0$, the critical field corresponding to the closing of the bulk quasiparticle gap is controlled by the SM-SC effective coupling $\lambda$ (see Sec.~\ref{sec:suppression}), while the zero-energy crossing of the in-gap ABS is correspondingly shifted to a higher Zeeman field.  The conductance map corresponding to the spectrum in Fig.~\ref{fig:EGdotABS}(c) is shown in Fig.~\ref{fig:EGdotABS}(d). Note the strong conductance peaks due to the dot-induced ABS, which represent the most prominent features of the conductance map. Unlike Fig.~\ref{fig:EGdotABS}(b), the closing of the bulk quasiparticle gap is visible, although the corresponding feature is particle-hole asymmetric due to the presence of finite dissipation. 



\begin{figure}[tbp]
\centering
\includegraphics[width=1.0\columnwidth]{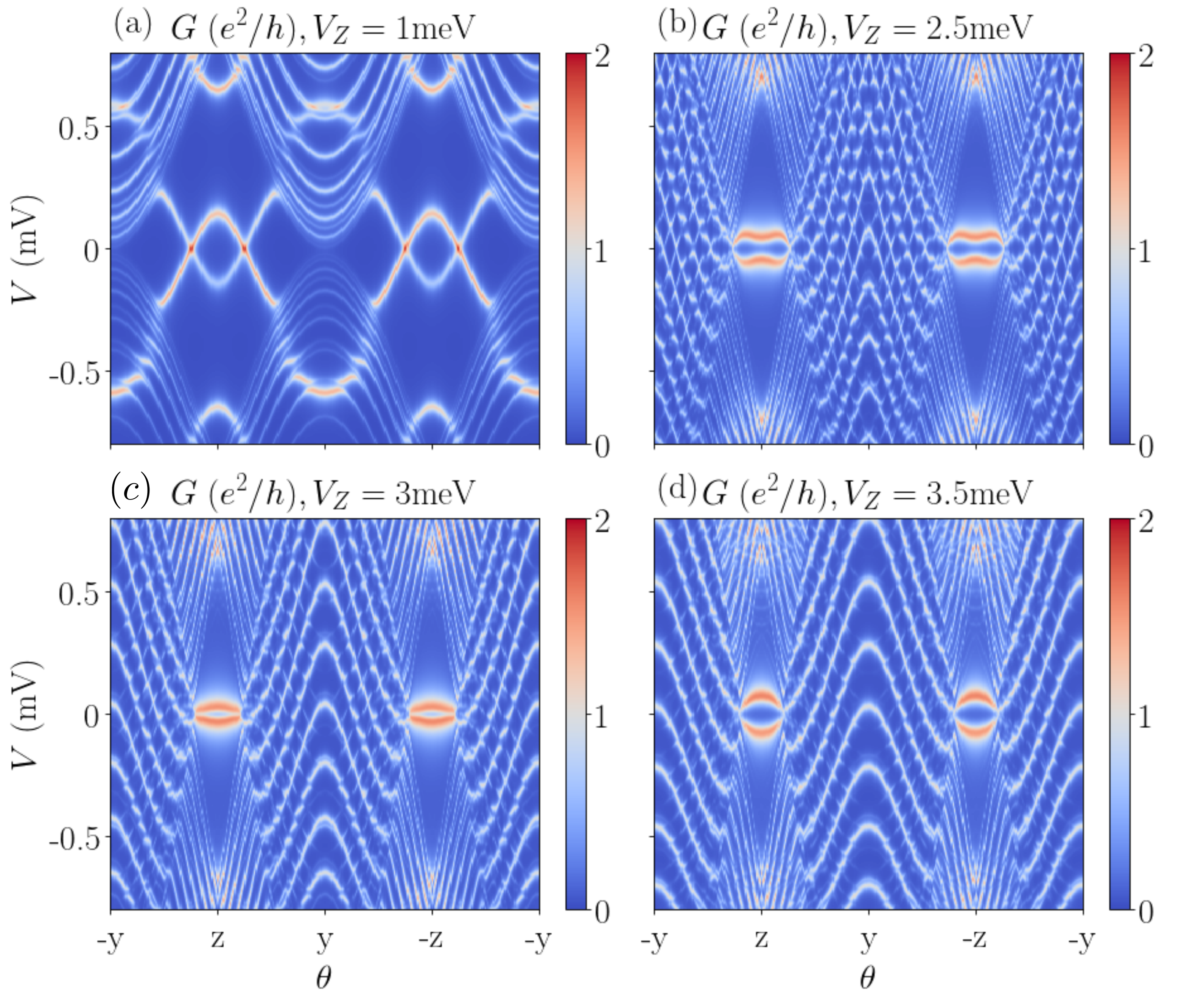}
\caption{Conductance map as a function of the magnetic field orientation (see Fig. \ref{fig:schematic} for the definition of the angle $\theta$) for a dot-nanowire system with parameters given in Fig. \ref{fig:EGdotABS}. The strength of the Zeeman field is fixed to be less than the critical value corresponding to the TQPT, so that all conductance plots are in the topologically trivial regime: (a) $V_Z=1$meV, (b) $V_Z=2.5$meV, (c) $V_Z=3$meV, (d) $V_Z=3.5$meV.}
\label{fig:RotationDot}
\end{figure}

The dependence of the conductance features on the orientation of the magnetic field within the plane perpendicular to the wire, which is given by the  angle $\theta$, is shown in Fig.~\ref{fig:RotationDot}. The signature of the low-energy ABS is clearly visible as a strong split conductance peak near $\theta = 0$ or $\pi$ (i.e., for field orientations approximately perpendicular to the spin-orbit field).  Outside this range of $\theta$, the conductance spectrum become gapless. Note that resolving the energy splitting of the near-zero conductance peak is a quantitative issue. Certain details regarding the quantum dot region (e.g., a different effective potential), finite temperature and dissipation, or proximity-induced renormalization effects could make this energy splitting unobservable, so that the feature associated with the dot-induced ABS becomes indistinguishable from a MZM-induced ZBCP. Both of them are effectively almost-zero-energy features in a finite wire with finite resolution although, strictly speaking, the MZM is precisely at zero energy (for a long wire) whereas the ABS is at a near-zero energy.

The natural question that arises in this context concerns the relationship between the main spectral features and the effective parameters describing the hybrid system. In particular, can one apply Eqs. (\ref{parameters1}-\ref{parameters3}) to estimate the effective parameters? First, we note that the presence of an in-gap state at $V_Z=0$ signals that corresponding states are partially localized outside the prioritized segment of the wire. Consequently, the system, which cannot be a pristine nanowire, is expected to contain some degree of inhomogeneity. In particular, the effective SM-SC coupling  characterizing the dot-induced ABS is different from (and smaller than) the effective SM-SC coupling characterizing states localized within the proximitized region. This ``bulk''  SM-SC coupling can be estimated using Eq.~\eqref{parameters1}, provided one can clearly identify the induced gap $\Delta_{\rm ind}$, which is larger than the energy of the in-gap ABS. Also, the (`bulk') effective $g$ factor can be estimated using  Eq.~\eqref{parameters2}, provided one can identify the critical field $B_c^y$ associated with the closing of the bulk gap. Note that $B_c^y$  is larger than the field corresponding to the zero crossing of the dot-induced ABS [see   Figs.~\ref{fig:RotationDot}(c) and (d)]. Finally, we note that one cannot estimate the chemical potential using Eq.~\eqref{parameters3} because there is no signature associated with the vanishing of the bulk gap, hence $B_c^x$ is unknown.

\begin{figure}[tbp]
\centering
\includegraphics[width=1.0\columnwidth]{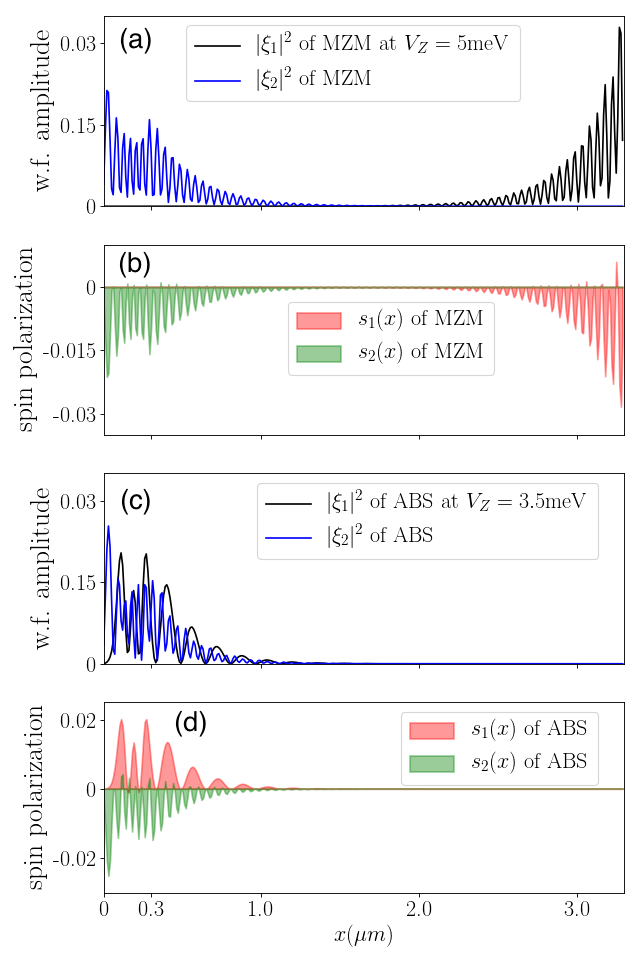}
\caption{Wave function amplitude and spin polarization (along the z-direction) for (a-b) MZMs and  (c-d) dot-induced ABS in a dot-nanowire system. In panels  (a) and (b), the strength of the Zeeman field is $V_Z=5~$meV (i.e., the nanowire is in the topological phase). Note the two spatially separated MZMs localized at the ends of the nanowire and the corresponding spin-down polarizations. In panels (c) and (d), the Zeeman field is $V_Z=3.5~$meV  (i.e., the nanowire is in the trivial phase). Note that the dot-induced ABS is localized at the left end of the wire (mostly in the dot region) and consists of two overlapping Majorana modes with opposite spin polarizations.}
\label{fig:wfDot}
\end{figure}

The above analysis has revealed similarities between the characteristic features of (quantum-dot-induced) almost-zero-energy ABSs emerging in the topologically trivial regime [e.g., the range $2 < V_Z < 4.5~$meV in Figs.~\ref{fig:EGdotABS}(a-b)], and the MZMs emerging  in the topological regime [$V_Z>4.5$meV  in Figs.~\ref{fig:EGdotABS}(a-b)]. While the signatures of these bound states in a local tunneling experiment are very similar, especially when the resolution of the  measurement is larger than the characteristic splitting amplitude, they have very different natures.   To better understand the fundamental difference between the dot-induced low-energy ABSs and the MZMs localized at the ends of the wire, we briefly discuss the real space and spin properties of their wave functions. Let $\hat{f}^{\dagger}$ be the creation operator for a generic BdG quasiparticle. We have
\begin{align}
\hat{f}^{\dagger} = \sum_{ \sigma, x } u_{\sigma}(x) \hat{c}^{\dagger}_{\sigma}(x) +v_{\sigma}(x) \hat{c}_{\sigma}(x),
\end{align}
where $\sigma$ and $x$ are the spin and position of the electron, respectively. One can always decompose the fermion operator $\hat{f}$ into two Majorana modes,
\begin{align}
\hat{\gamma}_1 &=  \hat{f}^{\dagger} + \hat{f} = \sum_{\sigma, x} \xi_{1\sigma}(x) \hat{c}^{\dagger}_{\sigma}(x) + \xi^*_{1\sigma}(x) \hat{c}_{\sigma}( x), \nn
\hat{\gamma}_2 &= i( \hat{f}^{\dagger} - \hat{f} ) = \sum_{\sigma, x} \xi_{2\sigma}(x) \hat{c}^{\dagger}_{\sigma}(x) + \xi^*_{2\sigma}(x) \hat{c}_{\sigma}( x), 
\end{align}
where $\xi_{1\sigma}$ ($\xi_{2\sigma}$) is the spin-$\sigma$ component of the wave function for the first (second) Majorana mode. To characterize the Majorana modes, it is convenient to analyze their real space properties, which are characterized by the wave function amplitude (for each Majorana mode  $i=1,2$) 
\begin{align}
|\xi_i|^2 = | \xi_{i \ua}(x) |^2 + | \xi_{i \da }(x) |^2.
\end{align}
as well as the spin properties, which are captured by the spin polarization along the z-direction defined as
\begin{align}
s_{i}(x) = \langle \sigma_z \rangle_{i} \equiv | \xi_{i \ua}(x) |^2 - | \xi_{i \da }(x) |^2.
\end{align}
The wave function amplitudes and the spin polarization characterizing the  MZMs and the dot-induced ABS emerging in a dot-nanowire system are shown in   Fig.~\ref{fig:wfDot}. The system parameters used in the numerical calculations are identical to those for Fig.~\ref{fig:EGdotABS}(a). For $V_Z=5~$meV, i.e., in the topological regime, the wave function of the lowest-energy BdG state is peaked near the ends of the wire and can be decomposed into two spatially separated MZMs, as shown in Fig.~\ref{fig:wfDot}(a). In this case, the two Majorana modes originate from the same spin-polarized subband (i.e., both have spin-down polarization), as shown in Fig.~\ref{fig:wfDot}(b). 
On the other hand, a near-zero-energy ABS that emerges below the critical field (e.g.,  $V_Z=3.5~$meV), consists of two overlapping Majorana modes localized  inside (and near) the quantum dot region, as shown in Fig.~\ref{fig:wfDot}(c). The two Majorana modes are spatially overlapped, while their spins are polarized in the opposite direction, as shown in Fig.~\ref{fig:wfDot}(d), indicating that the two Majoranas comprising the near-zero-energy ABS originate from opposite spin-polarized subbands. Note, however, that other potential profiles could generate low-energy ABSs with component Majorana modes associated with the same spin-polarized subband~\cite{Stanescu2018Illustrated}. 

\begin{figure}[tbp]
\centering
\includegraphics[width=1.0\columnwidth]{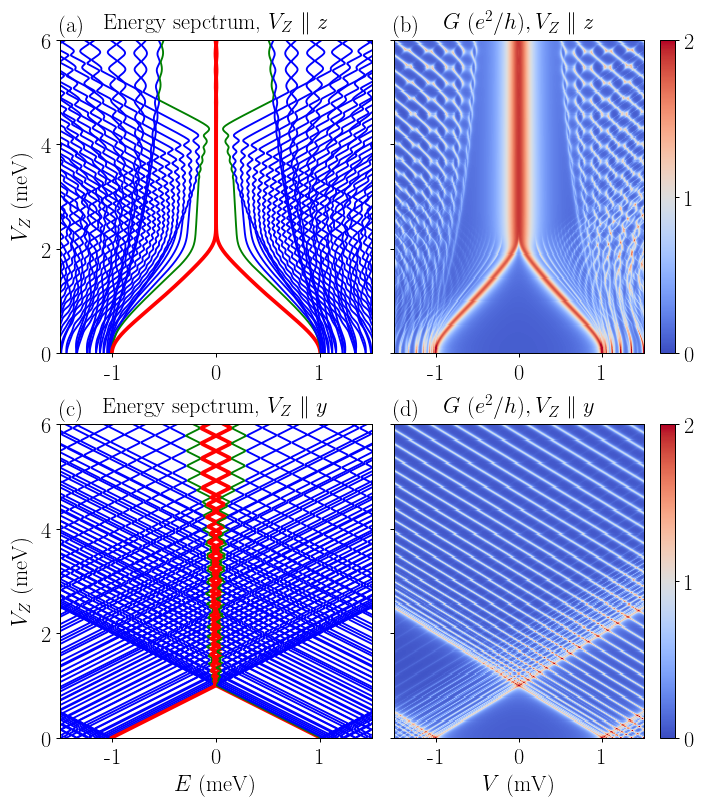}
\caption{Energy spectra and conductance maps for nanowire in the presence of chemical potential inhomogeneity. (a) and (b) are energy spectrum and conductance for Zeeman field along z-axis. Note that the bulk gap closes at $V_Z \simeq 2$meV, and stays small until the gap eventually reopens at at $V_Z \simeq 4.5~$meV. In the range of $2 < V_Z < 4.5~$meV, the nanowire is partially topological and the low-energy state is ABS. The corresponding conductance for such ABS is also $2e^2/h$ as shown in (b). (c) and (d) are energy spectrum and conductance for Zeeman field along y-axis. The SC gap closes as the Zeeman field increases, and the gap never reopens. There is no subgap ABS in the SC for small Zeeman field. }
\label{fig:EGinhomoABS}
\end{figure}

\subsection{Nanowire with smooth effective potential inhomogeneity}

In this subsection we examine another scenario that can give rise to (topologically trivial) low-energy ABS: the presence of smoothly varying effective potential inhomogeneities. As an example, we consider a linearly varying effective potential described by the BdG Hamiltonian 
\begin{align}
h_{\text{in}} = V_{\text{in}} ( 1 - x/L_{\text{wire}} )\tau_z , \quad 0 < x < L_{\text{wire}},
\label{eq:h_inhomo}
\end{align}
where $V_{\text{in}}$ is the amplitude of the effective potential variation and $L_{\text{wire}}$ is the length of the hybrid nanowire. The total Hamiltonian is the sum of  the terms contained in Eq.~\eqref{eq:1d_smsc} and the inhomogeneous potential described by Eq.~\eqref{eq:h_inhomo}. A schematic representation of this potential is given in Fig.~\ref{fig:Schematic_smooth}(b). In the numerical simulation, we choose $V_{\text{in}}=3~$meV and $L_{\text{wire}}=3~\mu$m.  

Fig.~\ref{fig:EGinhomoABS} shows the energy spectra and the conductance maps for a system with effective potential described by Eq.~\eqref{eq:h_inhomo} and Zeeman field oriented along two different directions. Figure~\ref{fig:EGinhomoABS}(a) represents the energy spectrum as a function of the field strength for a Zeeman field pointing along the z-axis. The lowest-energy mode (marked by the red line) collapses to zero energy way below the critical field associated with the whole nanowire becoming topologically nontrivial,  $V_Z \simeq 4.5~$meV, which corresponds to the reopening of the topological gap.  
Note that the quasiparticle gap (above the lowest lying mode) remains small over a finite range of Zeeman field ($2< V_Z < 4.5~$meV) before reopening, which  is a strong indication of partial topological superconductivity inside the nanowire. The low-energy ABS forming inside the nanowire in this range of Zeeman field generates a  ZBCP quantized at $2e^2/h$, making it completely indistinguishable from the MZM-induced ZBCP that emerges at larger fields, $V_Z > 4.5~$meV, [see Fig.~\ref{fig:EGinhomoABS}(b)].
Note that  in Fig.~\ref{fig:EGinhomoABS}(b) the structures associated with the bulk gap closing and reopening are completely suppressed as a result of the potential inhomogeneity reducing the amplitude of the corresponding wave functions at the left end of the system, hence reducing the coupling of these states to the tunnel probe. The energy spectrum and conductance map for a Zeeman field pointing along the y-axis (i.e., parallel to the spin-orbit effective field) are shown in Fig.~\ref{fig:EGinhomoABS}(c) and (d), respectively. Remarkably, the main features  are identical to those generated by a  pristine nanowire (i.e., a system with uniform effective potential), as shown in Fig.~\ref{fig:G_length}(b) and (e). Also,  unlike the nanowire coupled to an external quantum dot (see Fig.~\ref{fig:EGdotABS}), there is no subgap ABS at $V_Z=0$ and the closing of the bulk gap coincides with the ABS crossing zero energy. 

\begin{figure}[tbp]
\centering
\includegraphics[width=1.0\columnwidth]{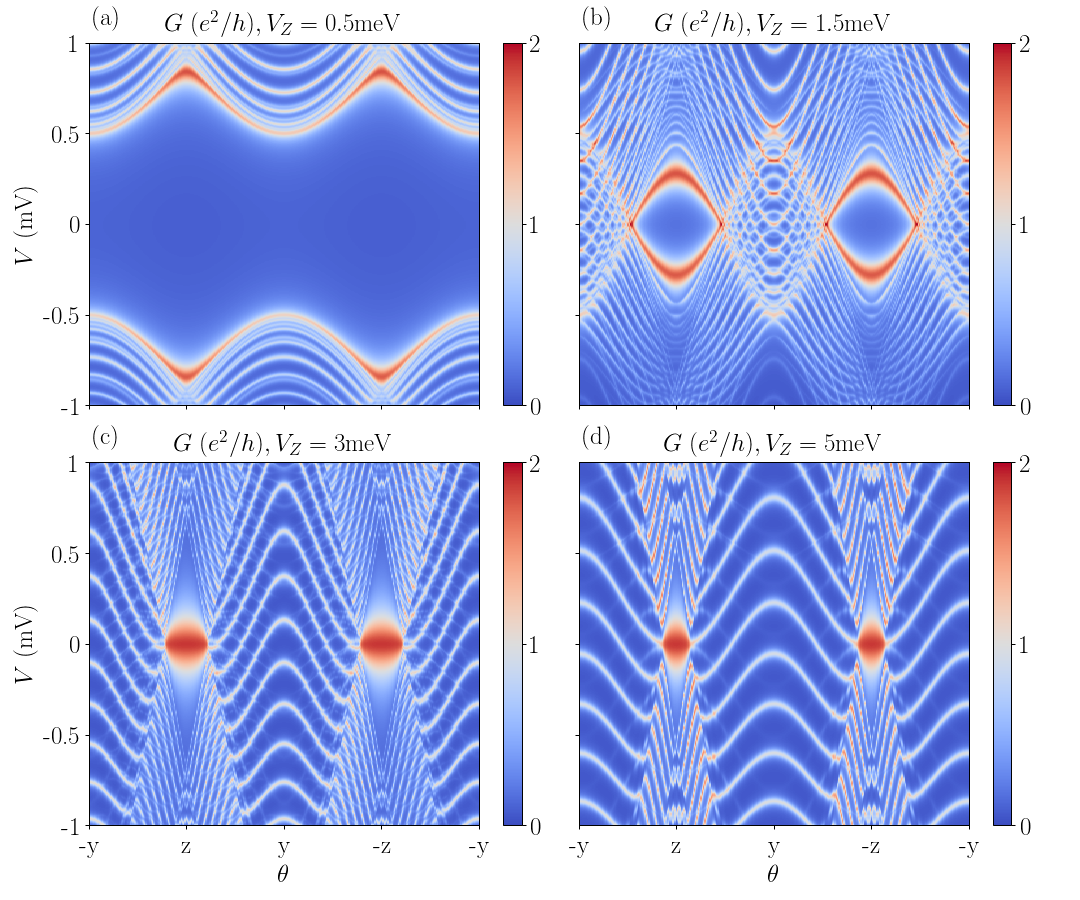}
\caption{Conductance map as a function of the magnetic field orientation (see Fig. \ref{fig:schematic} for the definition of the angle $\theta$) for a system with smooth inhomogeneous effective potential. The strength of the Zeeman field is fixed to: (a) $V_Z=0.5$meV, (b) $V_Z=1.5$meV, (c) $V_Z=3$meV, (d) $V_Z=5$meV.}
\label{fig:RotationInhomo}
\end{figure}

The dependence of tunnel conductance on the orientation of the Zeeman field in the plane perpendicular to the wire is shown in Fig.~\ref{fig:RotationInhomo}. When the field strength is smaller than the induced SC gap ($V_Z<\Delta_{\rm ind}=0.65~$meV), there is a hard gap in all directions [Fig.~\ref{fig:RotationInhomo}(a)]. The gap is maximum for fields oriented along the z-direction (i.e., perpendicular to the spin-orbit field) and minimum for fields oriented along the y-axis (i.e., parallel to the spin-orbit field).  When the strength of the Zeeman field becomes larger than the induced SC gap, the spectrum becomes gapless for $\theta\simeq \pm \pi/2$ (y-axis), while the conductance for $\theta \simeq 0/\pi$ (z-axis) is still gapped, as shown in Fig.~\ref{fig:RotationInhomo}(b). For large enough Zeeman fields (e.g., $V_Z=3~$meV), the smooth-potential-induced ABS collapses to zero energy and  the conductance map (as the function of $\theta$) becomes indistinguishable from a conduction map induced by MZMs, as evident from a comparison between Fig.~\ref{fig:RotationInhomo}(c) and (d). In both cases, as the Zeeman field rotates from a z-orientation towards the y-axis, the ABS- or MZM-induced ZBCP disappears and the spectrum becomes gapless.
  
\begin{figure}[tbp]
\centering
\includegraphics[width=1.0\columnwidth]{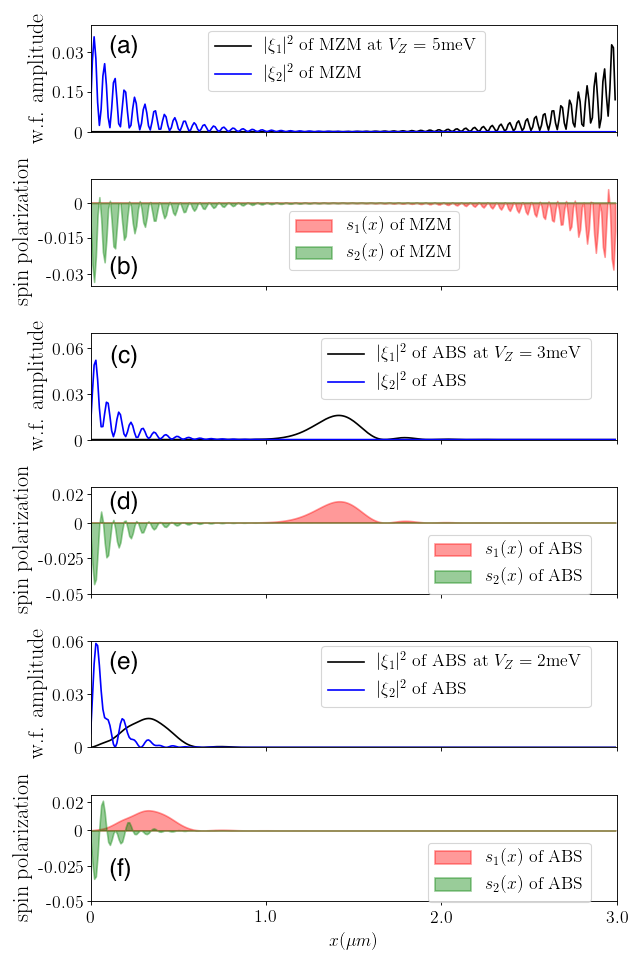}
\caption{Majorana wave function amplitudes and spin polarizations (along the z-direction) for (a-b) MZMs and  (c-d) the dot-induced ABS in a  system with smooth effective potential.  In panels (a) and (b) the strength of the Zeeman field is $V_Z=5~$meV and the entire nanowire is in the topological phase. The lowest energy modes  are two spatially separated MZMs localized at the ends of the system and originating from the same spin-split subband.  In panels (c) and (d) we have $V_Z=3~$meV and the nanowire is nominally in a topologically-trivial state. The lowest-energy state is a potential-induced ABS consisting of two well-separated Majorana modes associated with opposite spin polarized subbands, as shown by the corresponding spin polarizations. Panels  (e) and (f) correspond to  $V_Z=2.0~$meV (nanowire in the trivial phase). The component Majorana bound states of the low-energy ABS, which are only partially separated, have a significant overlap.}
\label{fig:wfInhomo}
\end{figure}

What is the impact of the phenomenology described in this section on the procedure for extracting the effective system parameters, as discussed in the context of Eqs.~(\ref{parameters1}-\ref{parameters3})?  First, we note that, unlike the dot-nanowire case, there are no obvious in-gap features to signal the presence of potential inhomogeneities. However, as before,  the effective SM-SC coupling can be estimated using Eq.~\eqref{parameters1} and the magnitude of the induced gap $\Delta_{\rm ind}$.  Also, the effective $g$-factor can be estimated using  Eq.~\eqref{parameters2} and the critical field $B_c^y$ associated with the closing of the bulk gap for a system with the field oriented along the spin-orbit field. Finally, we cannot estimate the chemical potential using Eq.\eqref{parameters3} because there is  no signature associated with the vanishing of the bulk gap at $B_c^x$. However, given the similarities between the conductance map in Fig.~\ref{fig:EGinhomoABS}(b) and the corresponding map for a pristine wire [e.g., Fig.~\ref{fig:G_length}(d)], which are virtually indistinguishable, one would wrongly estimate $B_c^x$ as the field associated with the collapse to zero-energy of the potential-induced ABS. The corresponding value of $\mu$ calculated using Eq.~\eqref{parameters3} will strongly underestimate the actual chemical potential. This type of systematic error can only be identified by further increasing the parameters space  (e.g., varying $\mu$ using potential gates) and detecting possible inconsistencies, or by performing nonlocal measurements (e.g., tunneling from both ends of the wire). This is a very serious issue requiring detailed experimental attention.

To gain further insight, we calculate the Majorana wave functions and the corresponding spin polarizations for the lowest-energy mode. The results are shown in  Fig.~\ref{fig:wfInhomo} for both the fully topological regime and the low-gap regime. In the fully topological case  ($V_Z > 4.5~$meV), the lowest energy mode consists of two MZMs localized at the opposite ends of the system and originating  from the same (spin-down) subband. The MZMs are weakly coupled to each other because of their large spatial separation, which is controlled by the length of the wire. By contrast, for the near-zero-energy ABS, the two component Majorana modes come from opposite spin-polarized subbands and have a field-dependent separation that is controlled (for a given field strength) by the slope of the effective potential. Note that Majorana mode with spin-up polarization, which  has a small characteristic wave vector (as indicated by the long-wavelength features), is localized at the potential domain wall separating regions with  empty and occupied spin-up subbands. Since inside the segment separating the two MZMs only one spin subband (the spin-down) is occupied, one can view the system as being ``locally topological''. Because of the (partial) separation of the component Majorana modes, only the left (spin-down) Majorana couples measurably to the tunnel probe, while the right (spin-up) Majorana remains ``invisible'' in a local tunneling measurement. Consequently, the corresponding ZBCP is quantized ($2e^2/h$) in both the fully topological (MZM)  and the low-gap (ABS) regimes. Figs.~\ref{fig:wfInhomo}(e) and (f) correspond to the situation of trivial phase with negligible topological superconductivity inside the nanowire ($V_Z=2.0~$meV). The component Majorana modes of the low-energy ABS are only partially separated, having a significant overlap. Thus, trivial ABS can mimic the behavior of topological MZMs in the rotating field measurements.



\section{Conclusion} \label{sec:conclusion}

In this paper we have investigated the anisotropic suppression of induced superconductivity in semiconductor-superconductor nanowires in the presence of magnetic fields with different orientations. The key questions are (i) what information about the parameters of the hybrid system can one extract from an angle-dependent conductance measurement, and (ii) is angle-dependent conductance spectroscopy capable of discriminating between signatures generated by topologically protected Majorana zero modes localized at the opposite ends of the wire and signatures produced by (topologically trivial) ABSs induced in nonhomogeneous systems? In this context, we have also examined various physical mechanisms potentially responsible for the suppression of the fine conductance structure apparent in simple model calculations, which can obscure key spectral feature and limit the information provided by tunnel conductance measurements.

First, we have examined the anisotropic suppression of induced superconductivity in pristine (homogeneous) nanowires. In this context, we have also investigated different physical mechanisms potentially responsible for the suppression of tunnel conductance features in the normal metal--superconductor junction, including the role of the effective wire length, the proximity coupling to the parent superconductor, finite temperature, finite dissipation, and multiband effects. We find that increasing the effective length of the wire leads to the reduction of the level spacing between bulk states, which makes the conductance maps smoother and structureless. We emphasize that a key parameter that controls the effective length of the system is the effective mass, which, in a one-dimensional semiconductor wire, is not simply a material-dependent constant, but a band-dependent property that is renormalized by the proximity coupling to the parent superconductor. Detailed microscopic calculations of the effective mass represent an outstanding theoretical task in this area. The proximity effect due to the coupling to the parent superconductor was considered within a self-energy approach. We find that above the parent SC gap the conductance is practically determined by the density of states of the parent superconductor and does not contain features associated with the semiconductor nanowire. Below the parent SC gap, the conductance features associated with bulk states become less well defined due to the proximity-induced low-energy renormalization. On the other hand, temperature and dissipation affect the resolution of conductance spectroscopy by broadening the spectral features. In addition, the presence of dissipation results in a particle-hole asymmetry of the tunneling conductance maps. Finally, we find that the relative weight of the subgap features associated with the top occupied band, including the Majorana-induced ZBCP and the ABS-induced gap closing signature, decreases with the number of occupied confinement-induced bands. Consequently, it is potentially more difficult to resolve the relevant low-energy features in nanowires with high occupancy. We emphasize that a combination of all these effects, rather than a single mechanism, is likely to be responsible  for the apparent discrepancy between experiment and simple model calculations with respect to the presence (absence) of conductance structures in theory (experiment).

Our investigation of magnetic field direction--dependent spectroscopy in pristine nanowires reveals a few well-defined, robust features that can be used to extract information about the parameters of the hybrid system. Particularly useful are the critical fields corresponding to the vanishing of the bulk gap: $B_c^x$, which is associated with the vanishing of the bulk gap at the TQPT in a system with a magnetic field oriented parallel to the wire (i.e., perpendicular to the effective spin-orbit field), and $B_c^y$, which is associated with the vanishing of the bulk gap in a system with a magnetic field oriented parallel to effective spin-orbit field. The direction of the effective spin-orbit field, which can be determined by performing an angle-dependent conductance measurement, is the direction corresponding to the minimum induced gap at low magnetic fields. Note that the closing of the bulk gap in pristine nanowires is signaled by intrinsic ABSs that emerge generically in systems with finite chemical potential and converge toward zero energy at the critical fields $B_c^x$ or $B_c^y$. In addition to these critical fields, a key observable is the induced gap at zero magnetic field, $\Delta_{\rm ind}$. We provide explicit equations that express the relationship between these observables and three key system parameters: the effective semiconductor-superconductor coupling, the effective $g$ factor, and the chemical potential. Using these relations, which hold if orbital and interband coupling effects are negligible, one can extract useful first-approximation estimates of these key parameters from conductance maps corresponding to different field orientations. We note that including orbital and interband coupling effects requires microscopic approaches that treat the semiconductor nanowire and the parent superconductor on an equal footing using, for example, multi-orbital two- or three-dimensional tight-binding models.

Second, we have investigated the anisotropic suppression of induced superconductivity in nonhomogeneous hybrid systems where near-zero-energy extrinsic nontopological ABSs mimic MZM properties. Our investigation focuses on two scenarios: (i) nanowires coupled to an (external) quantum dot, and (ii) nanowires with a smooth effective potential.  In both cases the inhomogeneity leads to the emergence of low-energy ABSs that collapse toward zero energy at field values lower than the critical field corresponding to the TQPT (i.e., in the nominally trivial phase). The signatures of these low-energy ABSs mimic the features induced by topologically protected Majorana zero modes localized at the ends of the wire. For the dot-wire system, we find that the presence of the external quantum dot is signaled by the emergence of in-gap states at zero magnetic field. These states are (partially) localized within the dot region and experience a reduced proximity coupling to the parent superconductor. On the other hand, we find that the conductance maps for the nanowire with smooth effective potential are virtually indistinguishable from the corresponding maps of a pristine, uniform nanowire. Thus, magnetic field direction-dependent tunneling spectroscopy by itself is unable to distinguish between a trivial ABS and topological MZM. In both cases it is still possible to extract the effective semiconductor-superconductor coupling and the effective $g$ factor based on a measurement with the magnetic field oriented parallel to the spin-orbit field, which provides the induced gap $\Delta_{\rm ind}$ and the critical field $B_c^y$. However, it is no longer possible to extract the chemical potential because there is no visible signature associated with the closing of the bulk gap at $B_c^x$.  While in pristine nanowires the closing of the gap is signaled by a feature induced by the intrinsic ABS, in the presence of inhomogeneities the ABS collapses to zero energy before the TQPT. By analyzing the wave function of the potential-induced extrinsic ABS, we find that the collapse to zero energy is associated with a partial separation of the Majorana components of the ABS. In addition, this partial separation results in one component Majorana being strongly coupled to the tunnel probe, while the other is practically ``invisible''. We conclude that one cannot clearly discriminate between MZM-induced features and features generated by low-energy ABSs emerging in nonhomogeneous systems based on angle-dependent conductance measurements. However, this type of measurement allows one to extract two key parameters, the effective SM-SC coupling and the effective $g$ factor, which, in turn, could facilitate the identification of possible discrepancies in our estimate of the chemical potential. Ultimately, verifying the consistency of the estimated chemical potential (which is crucial in order to clearly identify the topological regime) requires the expansion of the parameter space (e.g., by varying the gate-induced electrostatic potential) or the use of nonlocal data (e.g., conductance maps measured at the opposite ends of the system).


\begin{acknowledgements}
This work is supported by Laboratory for Physical Sciences and Microsoft. The authors thank Hao Zhang for helpful discussions on the experimental data of Ref.~\cite{Bommer2018SpinOrbit}. We acknowledge the support of University of Maryland supercomputing resources (http://hpcc.umd.edu). C.X.L. acknowledges the hospitality of the Kavli Institute of Theoretical Sciences at University of Chinese Academy of Sciences.
\end{acknowledgements}





\bibliography{BibMajorana}

\begin{thebibliography}{70}%
\makeatletter
\providecommand \@ifxundefined [1]{%
 \@ifx{#1\undefined}
}%
\providecommand \@ifnum [1]{%
 \ifnum #1\expandafter \@firstoftwo
 \else \expandafter \@secondoftwo
 \fi
}%
\providecommand \@ifx [1]{%
 \ifx #1\expandafter \@firstoftwo
 \else \expandafter \@secondoftwo
 \fi
}%
\providecommand \natexlab [1]{#1}%
\providecommand \enquote  [1]{``#1''}%
\providecommand \bibnamefont  [1]{#1}%
\providecommand \bibfnamefont [1]{#1}%
\providecommand \citenamefont [1]{#1}%
\providecommand \href@noop [0]{\@secondoftwo}%
\providecommand \href [0]{\begingroup \@sanitize@url \@href}%
\providecommand \@href[1]{\@@startlink{#1}\@@href}%
\providecommand \@@href[1]{\endgroup#1\@@endlink}%
\providecommand \@sanitize@url [0]{\catcode `\\12\catcode `\$12\catcode
  `\&12\catcode `\#12\catcode `\^12\catcode `\_12\catcode `\%12\relax}%
\providecommand \@@startlink[1]{}%
\providecommand \@@endlink[0]{}%
\providecommand \url  [0]{\begingroup\@sanitize@url \@url }%
\providecommand \@url [1]{\endgroup\@href {#1}{\urlprefix }}%
\providecommand \urlprefix  [0]{URL }%
\providecommand \Eprint [0]{\href }%
\providecommand \doibase [0]{http://dx.doi.org/}%
\providecommand \selectlanguage [0]{\@gobble}%
\providecommand \bibinfo  [0]{\@secondoftwo}%
\providecommand \bibfield  [0]{\@secondoftwo}%
\providecommand \translation [1]{[#1]}%
\providecommand \BibitemOpen [0]{}%
\providecommand \bibitemStop [0]{}%
\providecommand \bibitemNoStop [0]{.\EOS\space}%
\providecommand \EOS [0]{\spacefactor3000\relax}%
\providecommand \BibitemShut  [1]{\csname bibitem#1\endcsname}%
\let\auto@bib@innerbib\@empty
\bibitem [{\citenamefont {Nayak}\ \emph {et~al.}(2008)\citenamefont {Nayak},
  \citenamefont {Simon}, \citenamefont {Stern}, \citenamefont {Freedman},\ and\
  \citenamefont {Das~Sarma}}]{Nayak2008Non-Abelian}%
  \BibitemOpen
  \bibfield  {author} {\bibinfo {author} {\bibfnamefont {C.}~\bibnamefont
  {Nayak}}, \bibinfo {author} {\bibfnamefont {S.~H.}\ \bibnamefont {Simon}},
  \bibinfo {author} {\bibfnamefont {A.}~\bibnamefont {Stern}}, \bibinfo
  {author} {\bibfnamefont {M.}~\bibnamefont {Freedman}}, \ and\ \bibinfo
  {author} {\bibfnamefont {S.}~\bibnamefont {Das~Sarma}},\ }\href {\doibase
  10.1103/RevModPhys.80.1083} {\bibfield  {journal} {\bibinfo  {journal} {Rev.
  Mod. Phys.}\ }\textbf {\bibinfo {volume} {80}},\ \bibinfo {pages} {1083}
  (\bibinfo {year} {2008})}\BibitemShut {NoStop}%
\bibitem [{\citenamefont {Alicea}(2012)}]{Alicea2012New}%
  \BibitemOpen
  \bibfield  {author} {\bibinfo {author} {\bibfnamefont {J.}~\bibnamefont
  {Alicea}},\ }\href {http://stacks.iop.org/0034-4885/75/i=7/a=076501}
  {\bibfield  {journal} {\bibinfo  {journal} {Rep. Prog. Phys.}\ }\textbf
  {\bibinfo {volume} {75}},\ \bibinfo {pages} {076501} (\bibinfo {year}
  {2012})}\BibitemShut {NoStop}%
\bibitem [{\citenamefont {Leijnse}\ and\ \citenamefont
  {Flensberg}(2012)}]{Leijnse2012Introduction}%
  \BibitemOpen
  \bibfield  {author} {\bibinfo {author} {\bibfnamefont {M.}~\bibnamefont
  {Leijnse}}\ and\ \bibinfo {author} {\bibfnamefont {K.}~\bibnamefont
  {Flensberg}},\ }\href {http://stacks.iop.org/0268-1242/27/i=12/a=124003}
  {\bibfield  {journal} {\bibinfo  {journal} {Semicond. Sci. Technol.}\
  }\textbf {\bibinfo {volume} {27}},\ \bibinfo {pages} {124003} (\bibinfo
  {year} {2012})}\BibitemShut {NoStop}%
\bibitem [{\citenamefont {Beenakker}(2013)}]{Beenakker2013Search}%
  \BibitemOpen
  \bibfield  {author} {\bibinfo {author} {\bibfnamefont {C.}~\bibnamefont
  {Beenakker}},\ }\href {\doibase 10.1146/annurev-conmatphys-030212-184337}
  {\bibfield  {journal} {\bibinfo  {journal} {Annu. Rev. Condens. Matter
  Phys.}\ }\textbf {\bibinfo {volume} {4}},\ \bibinfo {pages} {113} (\bibinfo
  {year} {2013})}\BibitemShut {NoStop}%
\bibitem [{\citenamefont {Stanescu}\ and\ \citenamefont
  {Tewari}(2013)}]{Stanescu2013Majorana}%
  \BibitemOpen
  \bibfield  {author} {\bibinfo {author} {\bibfnamefont {T.~D.}\ \bibnamefont
  {Stanescu}}\ and\ \bibinfo {author} {\bibfnamefont {S.}~\bibnamefont
  {Tewari}},\ }\href
  {http://iopscience.iop.org/article/10.1088/0953-8984/25/23/233201/meta}
  {\bibfield  {journal} {\bibinfo  {journal} {J. Phys.: Condens. Matter}\
  }\textbf {\bibinfo {volume} {25}},\ \bibinfo {pages} {233201} (\bibinfo
  {year} {2013})}\BibitemShut {NoStop}%
\bibitem [{\citenamefont {Jiang}\ and\ \citenamefont
  {Wu}(2013)}]{Jiang2013Non}%
  \BibitemOpen
  \bibfield  {author} {\bibinfo {author} {\bibfnamefont {J.-H.}\ \bibnamefont
  {Jiang}}\ and\ \bibinfo {author} {\bibfnamefont {S.}~\bibnamefont {Wu}},\
  }\href {http://iopscience.iop.org/article/10.1088/0953-8984/25/5/055701/meta}
  {\bibfield  {journal} {\bibinfo  {journal} {J. Phys.: Condens. Matter}\
  }\textbf {\bibinfo {volume} {25}},\ \bibinfo {pages} {055701} (\bibinfo
  {year} {2013})}\BibitemShut {NoStop}%
\bibitem [{\citenamefont {Sarma}\ \emph {et~al.}(2015)\citenamefont {Sarma},
  \citenamefont {Freedman},\ and\ \citenamefont
  {Nayak}}]{DasSarma2015Majorana}%
  \BibitemOpen
  \bibfield  {author} {\bibinfo {author} {\bibfnamefont {S.~D.}\ \bibnamefont
  {Sarma}}, \bibinfo {author} {\bibfnamefont {M.}~\bibnamefont {Freedman}}, \
  and\ \bibinfo {author} {\bibfnamefont {C.}~\bibnamefont {Nayak}},\ }\href
  {http://dx.doi.org/10.1038/npjqi.2015.1} {\bibfield  {journal} {\bibinfo
  {journal} {Npj Quantum Information}\ }\textbf {\bibinfo {volume} {1}},\
  \bibinfo {pages} {15001 EP } (\bibinfo {year} {2015})}\BibitemShut {NoStop}%
\bibitem [{\citenamefont {Elliott}\ and\ \citenamefont
  {Franz}(2015)}]{Elliott2015Colloquium}%
  \BibitemOpen
  \bibfield  {author} {\bibinfo {author} {\bibfnamefont {S.~R.}\ \bibnamefont
  {Elliott}}\ and\ \bibinfo {author} {\bibfnamefont {M.}~\bibnamefont
  {Franz}},\ }\href {\doibase 10.1103/RevModPhys.87.137} {\bibfield  {journal}
  {\bibinfo  {journal} {Rev. Mod. Phys.}\ }\textbf {\bibinfo {volume} {87}},\
  \bibinfo {pages} {137} (\bibinfo {year} {2015})}\BibitemShut {NoStop}%
\bibitem [{\citenamefont {Sato}\ and\ \citenamefont
  {Fujimoto}(2016)}]{Sato2016Majorana}%
  \BibitemOpen
  \bibfield  {author} {\bibinfo {author} {\bibfnamefont {M.}~\bibnamefont
  {Sato}}\ and\ \bibinfo {author} {\bibfnamefont {S.}~\bibnamefont
  {Fujimoto}},\ }\href {\doibase 10.7566/JPSJ.85.072001} {\bibfield  {journal}
  {\bibinfo  {journal} {J. Phys. Soc. Jpn.}\ }\textbf {\bibinfo {volume}
  {85}},\ \bibinfo {pages} {072001} (\bibinfo {year} {2016})}\BibitemShut
  {NoStop}%
\bibitem [{\citenamefont {Sato}\ and\ \citenamefont
  {Ando}(2017)}]{Sato2017Topological}%
  \BibitemOpen
  \bibfield  {author} {\bibinfo {author} {\bibfnamefont {M.}~\bibnamefont
  {Sato}}\ and\ \bibinfo {author} {\bibfnamefont {Y.}~\bibnamefont {Ando}},\
  }\href {http://stacks.iop.org/0034-4885/80/i=7/a=076501} {\bibfield
  {journal} {\bibinfo  {journal} {Rep. Prog. Phys.}\ }\textbf {\bibinfo
  {volume} {80}},\ \bibinfo {pages} {076501} (\bibinfo {year}
  {2017})}\BibitemShut {NoStop}%
\bibitem [{\citenamefont {Aguado}(2017)}]{Aguado2017Majorana}%
  \BibitemOpen
  \bibfield  {author} {\bibinfo {author} {\bibfnamefont {R.}~\bibnamefont
  {Aguado}},\ }\href
  {https://www.sif.it/riviste/sif/ncr/econtents/2017/040/11/article/0}
  {\bibfield  {journal} {\bibinfo  {journal} {Riv. Nuovo Cimento}\ }\textbf
  {\bibinfo {volume} {40}},\ \bibinfo {pages} {523} (\bibinfo {year}
  {2017})}\BibitemShut {NoStop}%
\bibitem [{\citenamefont {Lutchyn}\ \emph {et~al.}(2018)\citenamefont
  {Lutchyn}, \citenamefont {Bakkers}, \citenamefont {Kouwenhoven},
  \citenamefont {Krogstrup}, \citenamefont {Marcus},\ and\ \citenamefont
  {Oreg}}]{Lutchyn2018Majorana}%
  \BibitemOpen
  \bibfield  {author} {\bibinfo {author} {\bibfnamefont {R.~M.}\ \bibnamefont
  {Lutchyn}}, \bibinfo {author} {\bibfnamefont {E.~P. A.~M.}\ \bibnamefont
  {Bakkers}}, \bibinfo {author} {\bibfnamefont {L.~P.}\ \bibnamefont
  {Kouwenhoven}}, \bibinfo {author} {\bibfnamefont {P.}~\bibnamefont
  {Krogstrup}}, \bibinfo {author} {\bibfnamefont {C.~M.}\ \bibnamefont
  {Marcus}}, \ and\ \bibinfo {author} {\bibfnamefont {Y.}~\bibnamefont
  {Oreg}},\ }\href {\doibase 10.1038/s41578-018-0003-1} {\bibfield  {journal}
  {\bibinfo  {journal} {Nat. Rev. Mater.}\ }\textbf {\bibinfo {volume} {3}},\
  \bibinfo {pages} {52} (\bibinfo {year} {2018})}\BibitemShut {NoStop}%
\bibitem [{\citenamefont {Sau}\ \emph {et~al.}(2010{\natexlab{a}})\citenamefont
  {Sau}, \citenamefont {Lutchyn}, \citenamefont {Tewari},\ and\ \citenamefont
  {Das~Sarma}}]{Sau2010Generic}%
  \BibitemOpen
  \bibfield  {author} {\bibinfo {author} {\bibfnamefont {J.~D.}\ \bibnamefont
  {Sau}}, \bibinfo {author} {\bibfnamefont {R.~M.}\ \bibnamefont {Lutchyn}},
  \bibinfo {author} {\bibfnamefont {S.}~\bibnamefont {Tewari}}, \ and\ \bibinfo
  {author} {\bibfnamefont {S.}~\bibnamefont {Das~Sarma}},\ }\href {\doibase
  10.1103/PhysRevLett.104.040502} {\bibfield  {journal} {\bibinfo  {journal}
  {Phys. Rev. Lett.}\ }\textbf {\bibinfo {volume} {104}},\ \bibinfo {pages}
  {040502} (\bibinfo {year} {2010}{\natexlab{a}})}\BibitemShut {NoStop}%
\bibitem [{\citenamefont {Lutchyn}\ \emph {et~al.}(2010)\citenamefont
  {Lutchyn}, \citenamefont {Sau},\ and\ \citenamefont
  {Das~Sarma}}]{Lutchyn2010Majorana}%
  \BibitemOpen
  \bibfield  {author} {\bibinfo {author} {\bibfnamefont {R.~M.}\ \bibnamefont
  {Lutchyn}}, \bibinfo {author} {\bibfnamefont {J.~D.}\ \bibnamefont {Sau}}, \
  and\ \bibinfo {author} {\bibfnamefont {S.}~\bibnamefont {Das~Sarma}},\ }\href
  {\doibase 10.1103/PhysRevLett.105.077001} {\bibfield  {journal} {\bibinfo
  {journal} {Phys. Rev. Lett.}\ }\textbf {\bibinfo {volume} {105}},\ \bibinfo
  {pages} {077001} (\bibinfo {year} {2010})}\BibitemShut {NoStop}%
\bibitem [{\citenamefont {Oreg}\ \emph {et~al.}(2010)\citenamefont {Oreg},
  \citenamefont {Refael},\ and\ \citenamefont {von Oppen}}]{Oreg2010Helical}%
  \BibitemOpen
  \bibfield  {author} {\bibinfo {author} {\bibfnamefont {Y.}~\bibnamefont
  {Oreg}}, \bibinfo {author} {\bibfnamefont {G.}~\bibnamefont {Refael}}, \ and\
  \bibinfo {author} {\bibfnamefont {F.}~\bibnamefont {von Oppen}},\ }\href
  {\doibase 10.1103/PhysRevLett.105.177002} {\bibfield  {journal} {\bibinfo
  {journal} {Phys. Rev. Lett.}\ }\textbf {\bibinfo {volume} {105}},\ \bibinfo
  {pages} {177002} (\bibinfo {year} {2010})}\BibitemShut {NoStop}%
\bibitem [{\citenamefont {Sau}\ \emph {et~al.}(2010{\natexlab{b}})\citenamefont
  {Sau}, \citenamefont {Tewari}, \citenamefont {Lutchyn}, \citenamefont
  {Stanescu},\ and\ \citenamefont {Das~Sarma}}]{Sau2010NonAbelian}%
  \BibitemOpen
  \bibfield  {author} {\bibinfo {author} {\bibfnamefont {J.~D.}\ \bibnamefont
  {Sau}}, \bibinfo {author} {\bibfnamefont {S.}~\bibnamefont {Tewari}},
  \bibinfo {author} {\bibfnamefont {R.~M.}\ \bibnamefont {Lutchyn}}, \bibinfo
  {author} {\bibfnamefont {T.~D.}\ \bibnamefont {Stanescu}}, \ and\ \bibinfo
  {author} {\bibfnamefont {S.}~\bibnamefont {Das~Sarma}},\ }\href {\doibase
  10.1103/PhysRevB.82.214509} {\bibfield  {journal} {\bibinfo  {journal} {Phys.
  Rev. B}\ }\textbf {\bibinfo {volume} {82}},\ \bibinfo {pages} {214509}
  (\bibinfo {year} {2010}{\natexlab{b}})}\BibitemShut {NoStop}%
\bibitem [{\citenamefont {Kitaev}(2001)}]{Kitaev2001Unpaired}%
  \BibitemOpen
  \bibfield  {author} {\bibinfo {author} {\bibfnamefont {A.~Y.}\ \bibnamefont
  {Kitaev}},\ }\href@noop {} {\bibfield  {journal} {\bibinfo  {journal}
  {Physics-Uspekhi}\ }\textbf {\bibinfo {volume} {44}},\ \bibinfo {pages} {131}
  (\bibinfo {year} {2001})}\BibitemShut {NoStop}%
\bibitem [{\citenamefont {Ivanov}(2001)}]{Ivanov2001NonAbelian}%
  \BibitemOpen
  \bibfield  {author} {\bibinfo {author} {\bibfnamefont {D.~A.}\ \bibnamefont
  {Ivanov}},\ }\href {\doibase 10.1103/PhysRevLett.86.268} {\bibfield
  {journal} {\bibinfo  {journal} {Phys. Rev. Lett.}\ }\textbf {\bibinfo
  {volume} {86}},\ \bibinfo {pages} {268} (\bibinfo {year} {2001})}\BibitemShut
  {NoStop}%
\bibitem [{\citenamefont {Alicea}\ \emph {et~al.}(2011)\citenamefont {Alicea},
  \citenamefont {Oreg}, \citenamefont {Refael}, \citenamefont {von Oppen},\
  and\ \citenamefont {Fisher}}]{Alicea2011NonAbelian}%
  \BibitemOpen
  \bibfield  {author} {\bibinfo {author} {\bibfnamefont {J.}~\bibnamefont
  {Alicea}}, \bibinfo {author} {\bibfnamefont {Y.}~\bibnamefont {Oreg}},
  \bibinfo {author} {\bibfnamefont {G.}~\bibnamefont {Refael}}, \bibinfo
  {author} {\bibfnamefont {F.}~\bibnamefont {von Oppen}}, \ and\ \bibinfo
  {author} {\bibfnamefont {M.~P.~A.}\ \bibnamefont {Fisher}},\ }\href
  {https://doi.org/10.1038/nphys1915} {\bibfield  {journal} {\bibinfo
  {journal} {Nature Physics}\ }\textbf {\bibinfo {volume} {7}},\ \bibinfo
  {pages} {412 EP } (\bibinfo {year} {2011})}\BibitemShut {NoStop}%
\bibitem [{\citenamefont {Aasen}\ \emph {et~al.}(2016)\citenamefont {Aasen},
  \citenamefont {Hell}, \citenamefont {Mishmash}, \citenamefont {Higginbotham},
  \citenamefont {Danon}, \citenamefont {Leijnse}, \citenamefont {Jespersen},
  \citenamefont {Folk}, \citenamefont {Marcus}, \citenamefont {Flensberg},\
  and\ \citenamefont {Alicea}}]{Aasen2016Milestones}%
  \BibitemOpen
  \bibfield  {author} {\bibinfo {author} {\bibfnamefont {D.}~\bibnamefont
  {Aasen}}, \bibinfo {author} {\bibfnamefont {M.}~\bibnamefont {Hell}},
  \bibinfo {author} {\bibfnamefont {R.~V.}\ \bibnamefont {Mishmash}}, \bibinfo
  {author} {\bibfnamefont {A.}~\bibnamefont {Higginbotham}}, \bibinfo {author}
  {\bibfnamefont {J.}~\bibnamefont {Danon}}, \bibinfo {author} {\bibfnamefont
  {M.}~\bibnamefont {Leijnse}}, \bibinfo {author} {\bibfnamefont {T.~S.}\
  \bibnamefont {Jespersen}}, \bibinfo {author} {\bibfnamefont {J.~A.}\
  \bibnamefont {Folk}}, \bibinfo {author} {\bibfnamefont {C.~M.}\ \bibnamefont
  {Marcus}}, \bibinfo {author} {\bibfnamefont {K.}~\bibnamefont {Flensberg}}, \
  and\ \bibinfo {author} {\bibfnamefont {J.}~\bibnamefont {Alicea}},\ }\href
  {\doibase 10.1103/PhysRevX.6.031016} {\bibfield  {journal} {\bibinfo
  {journal} {Phys. Rev. X}\ }\textbf {\bibinfo {volume} {6}},\ \bibinfo {pages}
  {031016} (\bibinfo {year} {2016})}\BibitemShut {NoStop}%
\bibitem [{\citenamefont {Vijay}\ and\ \citenamefont
  {Fu}(2016)}]{Vijay2016TeleportationBased}%
  \BibitemOpen
  \bibfield  {author} {\bibinfo {author} {\bibfnamefont {S.}~\bibnamefont
  {Vijay}}\ and\ \bibinfo {author} {\bibfnamefont {L.}~\bibnamefont {Fu}},\
  }\href {\doibase 10.1103/PhysRevB.94.235446} {\bibfield  {journal} {\bibinfo
  {journal} {Phys. Rev. B}\ }\textbf {\bibinfo {volume} {94}},\ \bibinfo
  {pages} {235446} (\bibinfo {year} {2016})}\BibitemShut {NoStop}%
\bibitem [{\citenamefont {Plugge}\ \emph {et~al.}(2017)\citenamefont {Plugge},
  \citenamefont {Rasmussen}, \citenamefont {Egger},\ and\ \citenamefont
  {Flensberg}}]{Plugge2017Majorana}%
  \BibitemOpen
  \bibfield  {author} {\bibinfo {author} {\bibfnamefont {S.}~\bibnamefont
  {Plugge}}, \bibinfo {author} {\bibfnamefont {A.}~\bibnamefont {Rasmussen}},
  \bibinfo {author} {\bibfnamefont {R.}~\bibnamefont {Egger}}, \ and\ \bibinfo
  {author} {\bibfnamefont {K.}~\bibnamefont {Flensberg}},\ }\href
  {http://stacks.iop.org/1367-2630/19/i=1/a=012001} {\bibfield  {journal}
  {\bibinfo  {journal} {New Journal of Physics}\ }\textbf {\bibinfo {volume}
  {19}},\ \bibinfo {pages} {012001} (\bibinfo {year} {2017})}\BibitemShut
  {NoStop}%
\bibitem [{\citenamefont {Karzig}\ \emph {et~al.}(2017)\citenamefont {Karzig},
  \citenamefont {Knapp}, \citenamefont {Lutchyn}, \citenamefont {Bonderson},
  \citenamefont {Hastings}, \citenamefont {Nayak}, \citenamefont {Alicea},
  \citenamefont {Flensberg}, \citenamefont {Plugge}, \citenamefont {Oreg},
  \citenamefont {Marcus},\ and\ \citenamefont {Freedman}}]{Karzig2017Scalable}%
  \BibitemOpen
  \bibfield  {author} {\bibinfo {author} {\bibfnamefont {T.}~\bibnamefont
  {Karzig}}, \bibinfo {author} {\bibfnamefont {C.}~\bibnamefont {Knapp}},
  \bibinfo {author} {\bibfnamefont {R.~M.}\ \bibnamefont {Lutchyn}}, \bibinfo
  {author} {\bibfnamefont {P.}~\bibnamefont {Bonderson}}, \bibinfo {author}
  {\bibfnamefont {M.~B.}\ \bibnamefont {Hastings}}, \bibinfo {author}
  {\bibfnamefont {C.}~\bibnamefont {Nayak}}, \bibinfo {author} {\bibfnamefont
  {J.}~\bibnamefont {Alicea}}, \bibinfo {author} {\bibfnamefont
  {K.}~\bibnamefont {Flensberg}}, \bibinfo {author} {\bibfnamefont
  {S.}~\bibnamefont {Plugge}}, \bibinfo {author} {\bibfnamefont
  {Y.}~\bibnamefont {Oreg}}, \bibinfo {author} {\bibfnamefont {C.~M.}\
  \bibnamefont {Marcus}}, \ and\ \bibinfo {author} {\bibfnamefont {M.~H.}\
  \bibnamefont {Freedman}},\ }\href {\doibase 10.1103/PhysRevB.95.235305}
  {\bibfield  {journal} {\bibinfo  {journal} {Phys. Rev. B}\ }\textbf {\bibinfo
  {volume} {95}},\ \bibinfo {pages} {235305} (\bibinfo {year}
  {2017})}\BibitemShut {NoStop}%
\bibitem [{\citenamefont {Sengupta}\ \emph {et~al.}(2001)\citenamefont
  {Sengupta}, \citenamefont {\ifmmode \check{Z}\else
  \v{Z}\fi{}uti\ifmmode~\acute{c}\else \'{c}\fi{}}, \citenamefont {Kwon},
  \citenamefont {Yakovenko},\ and\ \citenamefont
  {Das~Sarma}}]{Sengupta2001Midgap}%
  \BibitemOpen
  \bibfield  {author} {\bibinfo {author} {\bibfnamefont {K.}~\bibnamefont
  {Sengupta}}, \bibinfo {author} {\bibfnamefont {I.}~\bibnamefont {\ifmmode
  \check{Z}\else \v{Z}\fi{}uti\ifmmode~\acute{c}\else \'{c}\fi{}}}, \bibinfo
  {author} {\bibfnamefont {H.-J.}\ \bibnamefont {Kwon}}, \bibinfo {author}
  {\bibfnamefont {V.~M.}\ \bibnamefont {Yakovenko}}, \ and\ \bibinfo {author}
  {\bibfnamefont {S.}~\bibnamefont {Das~Sarma}},\ }\href {\doibase
  10.1103/PhysRevB.63.144531} {\bibfield  {journal} {\bibinfo  {journal} {Phys.
  Rev. B}\ }\textbf {\bibinfo {volume} {63}},\ \bibinfo {pages} {144531}
  (\bibinfo {year} {2001})}\BibitemShut {NoStop}%
\bibitem [{\citenamefont {Law}\ \emph {et~al.}(2009)\citenamefont {Law},
  \citenamefont {Lee},\ and\ \citenamefont {Ng}}]{Law2009Majorana}%
  \BibitemOpen
  \bibfield  {author} {\bibinfo {author} {\bibfnamefont {K.~T.}\ \bibnamefont
  {Law}}, \bibinfo {author} {\bibfnamefont {P.~A.}\ \bibnamefont {Lee}}, \ and\
  \bibinfo {author} {\bibfnamefont {T.~K.}\ \bibnamefont {Ng}},\ }\href
  {\doibase 10.1103/PhysRevLett.103.237001} {\bibfield  {journal} {\bibinfo
  {journal} {Phys. Rev. Lett.}\ }\textbf {\bibinfo {volume} {103}},\ \bibinfo
  {pages} {237001} (\bibinfo {year} {2009})}\BibitemShut {NoStop}%
\bibitem [{\citenamefont {Mourik}\ \emph {et~al.}(2012)\citenamefont {Mourik},
  \citenamefont {Zuo}, \citenamefont {Frolov}, \citenamefont {Plissard},
  \citenamefont {Bakkers},\ and\ \citenamefont
  {Kouwenhoven}}]{Mourik2012Signatures}%
  \BibitemOpen
  \bibfield  {author} {\bibinfo {author} {\bibfnamefont {V.}~\bibnamefont
  {Mourik}}, \bibinfo {author} {\bibfnamefont {K.}~\bibnamefont {Zuo}},
  \bibinfo {author} {\bibfnamefont {S.~M.}\ \bibnamefont {Frolov}}, \bibinfo
  {author} {\bibfnamefont {S.}~\bibnamefont {Plissard}}, \bibinfo {author}
  {\bibfnamefont {E.~P. A.~M.}\ \bibnamefont {Bakkers}}, \ and\ \bibinfo
  {author} {\bibfnamefont {L.~P.}\ \bibnamefont {Kouwenhoven}},\ }\href
  {\doibase 10.1126/science.1222360} {\bibfield  {journal} {\bibinfo  {journal}
  {Science}\ }\textbf {\bibinfo {volume} {336}},\ \bibinfo {pages} {1003}
  (\bibinfo {year} {2012})}\BibitemShut {NoStop}%
\bibitem [{\citenamefont {Das}\ \emph {et~al.}(2012)\citenamefont {Das},
  \citenamefont {Ronen}, \citenamefont {Most}, \citenamefont {Oreg},
  \citenamefont {Heiblum},\ and\ \citenamefont {Shtrikman}}]{Das2012Zero}%
  \BibitemOpen
  \bibfield  {author} {\bibinfo {author} {\bibfnamefont {A.}~\bibnamefont
  {Das}}, \bibinfo {author} {\bibfnamefont {Y.}~\bibnamefont {Ronen}}, \bibinfo
  {author} {\bibfnamefont {Y.}~\bibnamefont {Most}}, \bibinfo {author}
  {\bibfnamefont {Y.}~\bibnamefont {Oreg}}, \bibinfo {author} {\bibfnamefont
  {M.}~\bibnamefont {Heiblum}}, \ and\ \bibinfo {author} {\bibfnamefont
  {H.}~\bibnamefont {Shtrikman}},\ }\href {http://dx.doi.org/10.1038/nphys2479}
  {\bibfield  {journal} {\bibinfo  {journal} {Nat. Phys.}\ }\textbf {\bibinfo
  {volume} {8}},\ \bibinfo {pages} {887} (\bibinfo {year} {2012})}\BibitemShut
  {NoStop}%
\bibitem [{\citenamefont {Deng}\ \emph {et~al.}(2012)\citenamefont {Deng},
  \citenamefont {Yu}, \citenamefont {Huang}, \citenamefont {Larsson},
  \citenamefont {Caroff},\ and\ \citenamefont {Xu}}]{Deng2012Anomalous}%
  \BibitemOpen
  \bibfield  {author} {\bibinfo {author} {\bibfnamefont {M.~T.}\ \bibnamefont
  {Deng}}, \bibinfo {author} {\bibfnamefont {C.~L.}\ \bibnamefont {Yu}},
  \bibinfo {author} {\bibfnamefont {G.~Y.}\ \bibnamefont {Huang}}, \bibinfo
  {author} {\bibfnamefont {M.}~\bibnamefont {Larsson}}, \bibinfo {author}
  {\bibfnamefont {P.}~\bibnamefont {Caroff}}, \ and\ \bibinfo {author}
  {\bibfnamefont {H.~Q.}\ \bibnamefont {Xu}},\ }\href {\doibase
  10.1021/nl303758w} {\bibfield  {journal} {\bibinfo  {journal} {Nano Lett.}\
  }\textbf {\bibinfo {volume} {12}},\ \bibinfo {pages} {6414} (\bibinfo {year}
  {2012})}\BibitemShut {NoStop}%
\bibitem [{\citenamefont {Churchill}\ \emph {et~al.}(2013)\citenamefont
  {Churchill}, \citenamefont {Fatemi}, \citenamefont {Grove-Rasmussen},
  \citenamefont {Deng}, \citenamefont {Caroff}, \citenamefont {Xu},\ and\
  \citenamefont {Marcus}}]{Churchill2013Superconductor}%
  \BibitemOpen
  \bibfield  {author} {\bibinfo {author} {\bibfnamefont {H.~O.~H.}\
  \bibnamefont {Churchill}}, \bibinfo {author} {\bibfnamefont {V.}~\bibnamefont
  {Fatemi}}, \bibinfo {author} {\bibfnamefont {K.}~\bibnamefont
  {Grove-Rasmussen}}, \bibinfo {author} {\bibfnamefont {M.~T.}\ \bibnamefont
  {Deng}}, \bibinfo {author} {\bibfnamefont {P.}~\bibnamefont {Caroff}},
  \bibinfo {author} {\bibfnamefont {H.~Q.}\ \bibnamefont {Xu}}, \ and\ \bibinfo
  {author} {\bibfnamefont {C.~M.}\ \bibnamefont {Marcus}},\ }\href {\doibase
  10.1103/PhysRevB.87.241401} {\bibfield  {journal} {\bibinfo  {journal} {Phys.
  Rev. B}\ }\textbf {\bibinfo {volume} {87}},\ \bibinfo {pages} {241401}
  (\bibinfo {year} {2013})}\BibitemShut {NoStop}%
\bibitem [{\citenamefont {Finck}\ \emph {et~al.}(2013)\citenamefont {Finck},
  \citenamefont {Van~Harlingen}, \citenamefont {Mohseni}, \citenamefont
  {Jung},\ and\ \citenamefont {Li}}]{Finck2013Anomalous}%
  \BibitemOpen
  \bibfield  {author} {\bibinfo {author} {\bibfnamefont {A.~D.~K.}\
  \bibnamefont {Finck}}, \bibinfo {author} {\bibfnamefont {D.~J.}\ \bibnamefont
  {Van~Harlingen}}, \bibinfo {author} {\bibfnamefont {P.~K.}\ \bibnamefont
  {Mohseni}}, \bibinfo {author} {\bibfnamefont {K.}~\bibnamefont {Jung}}, \
  and\ \bibinfo {author} {\bibfnamefont {X.}~\bibnamefont {Li}},\ }\href
  {\doibase 10.1103/PhysRevLett.110.126406} {\bibfield  {journal} {\bibinfo
  {journal} {Phys. Rev. Lett.}\ }\textbf {\bibinfo {volume} {110}},\ \bibinfo
  {pages} {126406} (\bibinfo {year} {2013})}\BibitemShut {NoStop}%
\bibitem [{\citenamefont {Albrecht}\ \emph {et~al.}(2016)\citenamefont
  {Albrecht}, \citenamefont {Higginbotham}, \citenamefont {Madsen},
  \citenamefont {Kuemmeth}, \citenamefont {Jespersen}, \citenamefont
  {Nyg{\aa}rd}, \citenamefont {Krogstrup},\ and\ \citenamefont
  {Marcus}}]{Albrecht2016Exponential}%
  \BibitemOpen
  \bibfield  {author} {\bibinfo {author} {\bibfnamefont {S.}~\bibnamefont
  {Albrecht}}, \bibinfo {author} {\bibfnamefont {A.}~\bibnamefont
  {Higginbotham}}, \bibinfo {author} {\bibfnamefont {M.}~\bibnamefont
  {Madsen}}, \bibinfo {author} {\bibfnamefont {F.}~\bibnamefont {Kuemmeth}},
  \bibinfo {author} {\bibfnamefont {T.}~\bibnamefont {Jespersen}}, \bibinfo
  {author} {\bibfnamefont {J.}~\bibnamefont {Nyg{\aa}rd}}, \bibinfo {author}
  {\bibfnamefont {P.}~\bibnamefont {Krogstrup}}, \ and\ \bibinfo {author}
  {\bibfnamefont {C.}~\bibnamefont {Marcus}},\ }\href
  {http://dx.doi.org/10.1038/nature17162} {\bibfield  {journal} {\bibinfo
  {journal} {Nature}\ }\textbf {\bibinfo {volume} {531}},\ \bibinfo {pages}
  {206} (\bibinfo {year} {2016})}\BibitemShut {NoStop}%
\bibitem [{\citenamefont {Chen}\ \emph {et~al.}(2017)\citenamefont {Chen},
  \citenamefont {Yu}, \citenamefont {Stenger}, \citenamefont {Hocevar},
  \citenamefont {Car}, \citenamefont {Plissard}, \citenamefont {Bakkers},
  \citenamefont {Stanescu},\ and\ \citenamefont
  {Frolov}}]{Chen2017Experimental}%
  \BibitemOpen
  \bibfield  {author} {\bibinfo {author} {\bibfnamefont {J.}~\bibnamefont
  {Chen}}, \bibinfo {author} {\bibfnamefont {P.}~\bibnamefont {Yu}}, \bibinfo
  {author} {\bibfnamefont {J.}~\bibnamefont {Stenger}}, \bibinfo {author}
  {\bibfnamefont {M.}~\bibnamefont {Hocevar}}, \bibinfo {author} {\bibfnamefont
  {D.}~\bibnamefont {Car}}, \bibinfo {author} {\bibfnamefont {S.~R.}\
  \bibnamefont {Plissard}}, \bibinfo {author} {\bibfnamefont {E.~P. A.~M.}\
  \bibnamefont {Bakkers}}, \bibinfo {author} {\bibfnamefont {T.~D.}\
  \bibnamefont {Stanescu}}, \ and\ \bibinfo {author} {\bibfnamefont {S.~M.}\
  \bibnamefont {Frolov}},\ }\href {\doibase 10.1126/sciadv.1701476} {\bibfield
  {journal} {\bibinfo  {journal} {Science Advances}\ }\textbf {\bibinfo
  {volume} {3}} (\bibinfo {year} {2017}),\ 10.1126/sciadv.1701476}\BibitemShut
  {NoStop}%
\bibitem [{\citenamefont {Deng}\ \emph {et~al.}(2016)\citenamefont {Deng},
  \citenamefont {Vaitiekenas}, \citenamefont {Hansen}, \citenamefont {Danon},
  \citenamefont {Leijnse}, \citenamefont {Flensberg}, \citenamefont
  {Nyg{\aa}rd}, \citenamefont {Krogstrup},\ and\ \citenamefont
  {Marcus}}]{Deng2016Majorana}%
  \BibitemOpen
  \bibfield  {author} {\bibinfo {author} {\bibfnamefont {M.~T.}\ \bibnamefont
  {Deng}}, \bibinfo {author} {\bibfnamefont {S.}~\bibnamefont {Vaitiekenas}},
  \bibinfo {author} {\bibfnamefont {E.~B.}\ \bibnamefont {Hansen}}, \bibinfo
  {author} {\bibfnamefont {J.}~\bibnamefont {Danon}}, \bibinfo {author}
  {\bibfnamefont {M.}~\bibnamefont {Leijnse}}, \bibinfo {author} {\bibfnamefont
  {K.}~\bibnamefont {Flensberg}}, \bibinfo {author} {\bibfnamefont
  {J.}~\bibnamefont {Nyg{\aa}rd}}, \bibinfo {author} {\bibfnamefont
  {P.}~\bibnamefont {Krogstrup}}, \ and\ \bibinfo {author} {\bibfnamefont
  {C.~M.}\ \bibnamefont {Marcus}},\ }\href {\doibase 10.1126/science.aaf3961}
  {\bibfield  {journal} {\bibinfo  {journal} {Science}\ }\textbf {\bibinfo
  {volume} {354}},\ \bibinfo {pages} {1557} (\bibinfo {year}
  {2016})}\BibitemShut {NoStop}%
\bibitem [{\citenamefont {Zhang}\ \emph {et~al.}(2017)\citenamefont {Zhang},
  \citenamefont {G{\"u}l}, \citenamefont {Conesa-Boj}, \citenamefont {Nowak},
  \citenamefont {Wimmer}, \citenamefont {Zuo}, \citenamefont {Mourik},
  \citenamefont {de~Vries}, \citenamefont {van Veen}, \citenamefont {de~Moor},
  \citenamefont {Bommer}, \citenamefont {van Woerkom}, \citenamefont {Car},
  \citenamefont {Plissard}, \citenamefont {Bakkers}, \citenamefont
  {Quintero-P{\'e}rez}, \citenamefont {Cassidy}, \citenamefont {Koelling},
  \citenamefont {Goswami}, \citenamefont {Watanabe}, \citenamefont
  {Taniguchi},\ and\ \citenamefont {Kouwenhoven}}]{Zhang2017Ballistic}%
  \BibitemOpen
  \bibfield  {author} {\bibinfo {author} {\bibfnamefont {H.}~\bibnamefont
  {Zhang}}, \bibinfo {author} {\bibfnamefont {{\"O}.}~\bibnamefont {G{\"u}l}},
  \bibinfo {author} {\bibfnamefont {S.}~\bibnamefont {Conesa-Boj}}, \bibinfo
  {author} {\bibfnamefont {M.}~\bibnamefont {Nowak}}, \bibinfo {author}
  {\bibfnamefont {M.}~\bibnamefont {Wimmer}}, \bibinfo {author} {\bibfnamefont
  {K.}~\bibnamefont {Zuo}}, \bibinfo {author} {\bibfnamefont {V.}~\bibnamefont
  {Mourik}}, \bibinfo {author} {\bibfnamefont {F.~K.}\ \bibnamefont
  {de~Vries}}, \bibinfo {author} {\bibfnamefont {J.}~\bibnamefont {van Veen}},
  \bibinfo {author} {\bibfnamefont {M.~W.~A.}\ \bibnamefont {de~Moor}},
  \bibinfo {author} {\bibfnamefont {J.~D.~S.}\ \bibnamefont {Bommer}}, \bibinfo
  {author} {\bibfnamefont {D.~J.}\ \bibnamefont {van Woerkom}}, \bibinfo
  {author} {\bibfnamefont {D.}~\bibnamefont {Car}}, \bibinfo {author}
  {\bibfnamefont {S.~R.}\ \bibnamefont {Plissard}}, \bibinfo {author}
  {\bibfnamefont {E.~P. A.~M.}\ \bibnamefont {Bakkers}}, \bibinfo {author}
  {\bibfnamefont {M.}~\bibnamefont {Quintero-P{\'e}rez}}, \bibinfo {author}
  {\bibfnamefont {M.~C.}\ \bibnamefont {Cassidy}}, \bibinfo {author}
  {\bibfnamefont {S.}~\bibnamefont {Koelling}}, \bibinfo {author}
  {\bibfnamefont {S.}~\bibnamefont {Goswami}}, \bibinfo {author} {\bibfnamefont
  {K.}~\bibnamefont {Watanabe}}, \bibinfo {author} {\bibfnamefont
  {T.}~\bibnamefont {Taniguchi}}, \ and\ \bibinfo {author} {\bibfnamefont
  {L.~P.}\ \bibnamefont {Kouwenhoven}},\ }\href
  {http://dx.doi.org/10.1038/ncomms16025} {\bibfield  {journal} {\bibinfo
  {journal} {Nature Communications}\ }\textbf {\bibinfo {volume} {8}},\
  \bibinfo {pages} {16025 EP } (\bibinfo {year} {2017})}\BibitemShut {NoStop}%
\bibitem [{\citenamefont {G{\"u}l}\ \emph {et~al.}(2018)\citenamefont
  {G{\"u}l}, \citenamefont {Zhang}, \citenamefont {Bommer}, \citenamefont
  {de~Moor}, \citenamefont {Car}, \citenamefont {Plissard}, \citenamefont
  {Bakkers}, \citenamefont {Geresdi}, \citenamefont {Watanabe}, \citenamefont
  {Taniguchi},\ and\ \citenamefont {Kouwenhoven}}]{Gul2018Ballistic}%
  \BibitemOpen
  \bibfield  {author} {\bibinfo {author} {\bibfnamefont {{\"O}.}~\bibnamefont
  {G{\"u}l}}, \bibinfo {author} {\bibfnamefont {H.}~\bibnamefont {Zhang}},
  \bibinfo {author} {\bibfnamefont {J.~D.~S.}\ \bibnamefont {Bommer}}, \bibinfo
  {author} {\bibfnamefont {M.~W.~A.}\ \bibnamefont {de~Moor}}, \bibinfo
  {author} {\bibfnamefont {D.}~\bibnamefont {Car}}, \bibinfo {author}
  {\bibfnamefont {S.~R.}\ \bibnamefont {Plissard}}, \bibinfo {author}
  {\bibfnamefont {E.~P. A.~M.}\ \bibnamefont {Bakkers}}, \bibinfo {author}
  {\bibfnamefont {A.}~\bibnamefont {Geresdi}}, \bibinfo {author} {\bibfnamefont
  {K.}~\bibnamefont {Watanabe}}, \bibinfo {author} {\bibfnamefont
  {T.}~\bibnamefont {Taniguchi}}, \ and\ \bibinfo {author} {\bibfnamefont
  {L.~P.}\ \bibnamefont {Kouwenhoven}},\ }\href
  {https://www.nature.com/articles/s41565-017-0032-8} {\bibfield  {journal}
  {\bibinfo  {journal} {Nat. Nanotechnol.}\ }\textbf {\bibinfo {volume} {13}},\
  \bibinfo {pages} {192} (\bibinfo {year} {2018})}\BibitemShut {NoStop}%
\bibitem [{\citenamefont {Nichele}\ \emph {et~al.}(2017)\citenamefont
  {Nichele}, \citenamefont {Drachmann}, \citenamefont {Whiticar}, \citenamefont
  {O'Farrell}, \citenamefont {Suominen}, \citenamefont {Fornieri},
  \citenamefont {Wang}, \citenamefont {Gardner}, \citenamefont {Thomas},
  \citenamefont {Hatke}, \citenamefont {Krogstrup}, \citenamefont {Manfra},
  \citenamefont {Flensberg},\ and\ \citenamefont
  {Marcus}}]{Nichele2017Scaling}%
  \BibitemOpen
  \bibfield  {author} {\bibinfo {author} {\bibfnamefont {F.}~\bibnamefont
  {Nichele}}, \bibinfo {author} {\bibfnamefont {A.~C.~C.}\ \bibnamefont
  {Drachmann}}, \bibinfo {author} {\bibfnamefont {A.~M.}\ \bibnamefont
  {Whiticar}}, \bibinfo {author} {\bibfnamefont {E.~C.~T.}\ \bibnamefont
  {O'Farrell}}, \bibinfo {author} {\bibfnamefont {H.~J.}\ \bibnamefont
  {Suominen}}, \bibinfo {author} {\bibfnamefont {A.}~\bibnamefont {Fornieri}},
  \bibinfo {author} {\bibfnamefont {T.}~\bibnamefont {Wang}}, \bibinfo {author}
  {\bibfnamefont {G.~C.}\ \bibnamefont {Gardner}}, \bibinfo {author}
  {\bibfnamefont {C.}~\bibnamefont {Thomas}}, \bibinfo {author} {\bibfnamefont
  {A.~T.}\ \bibnamefont {Hatke}}, \bibinfo {author} {\bibfnamefont
  {P.}~\bibnamefont {Krogstrup}}, \bibinfo {author} {\bibfnamefont {M.~J.}\
  \bibnamefont {Manfra}}, \bibinfo {author} {\bibfnamefont {K.}~\bibnamefont
  {Flensberg}}, \ and\ \bibinfo {author} {\bibfnamefont {C.~M.}\ \bibnamefont
  {Marcus}},\ }\href {\doibase 10.1103/PhysRevLett.119.136803} {\bibfield
  {journal} {\bibinfo  {journal} {Phys. Rev. Lett.}\ }\textbf {\bibinfo
  {volume} {119}},\ \bibinfo {pages} {136803} (\bibinfo {year}
  {2017})}\BibitemShut {NoStop}%
\bibitem [{\citenamefont {Zhang}\ \emph {et~al.}(2018)\citenamefont {Zhang},
  \citenamefont {Liu}, \citenamefont {Gazibegovic}, \citenamefont {Xu},
  \citenamefont {Logan}, \citenamefont {Wang}, \citenamefont {Van~Loo},
  \citenamefont {Bommer}, \citenamefont {De~Moor}, \citenamefont {Car} \emph
  {et~al.}}]{Zhang2017Quantized}%
  \BibitemOpen
  \bibfield  {author} {\bibinfo {author} {\bibfnamefont {H.}~\bibnamefont
  {Zhang}}, \bibinfo {author} {\bibfnamefont {C.-X.}\ \bibnamefont {Liu}},
  \bibinfo {author} {\bibfnamefont {S.}~\bibnamefont {Gazibegovic}}, \bibinfo
  {author} {\bibfnamefont {D.}~\bibnamefont {Xu}}, \bibinfo {author}
  {\bibfnamefont {J.~A.}\ \bibnamefont {Logan}}, \bibinfo {author}
  {\bibfnamefont {G.}~\bibnamefont {Wang}}, \bibinfo {author} {\bibfnamefont
  {N.}~\bibnamefont {Van~Loo}}, \bibinfo {author} {\bibfnamefont {J.~D.}\
  \bibnamefont {Bommer}}, \bibinfo {author} {\bibfnamefont {M.~W.}\
  \bibnamefont {De~Moor}}, \bibinfo {author} {\bibfnamefont {D.}~\bibnamefont
  {Car}},  \emph {et~al.},\ }\href {\doibase 10.1038/nature26142} {\bibfield
  {journal} {\bibinfo  {journal} {Nature}\ }\textbf {\bibinfo {volume} {556}},\
  \bibinfo {pages} {74} (\bibinfo {year} {2018})}\BibitemShut {NoStop}%
\bibitem [{\citenamefont {Kells}\ \emph {et~al.}(2012)\citenamefont {Kells},
  \citenamefont {Meidan},\ and\ \citenamefont {Brouwer}}]{Kells2012Near}%
  \BibitemOpen
  \bibfield  {author} {\bibinfo {author} {\bibfnamefont {G.}~\bibnamefont
  {Kells}}, \bibinfo {author} {\bibfnamefont {D.}~\bibnamefont {Meidan}}, \
  and\ \bibinfo {author} {\bibfnamefont {P.~W.}\ \bibnamefont {Brouwer}},\
  }\href {\doibase 10.1103/PhysRevB.86.100503} {\bibfield  {journal} {\bibinfo
  {journal} {Phys. Rev. B}\ }\textbf {\bibinfo {volume} {86}},\ \bibinfo
  {pages} {100503} (\bibinfo {year} {2012})}\BibitemShut {NoStop}%
\bibitem [{\citenamefont {Prada}\ \emph {et~al.}(2012)\citenamefont {Prada},
  \citenamefont {San-Jose},\ and\ \citenamefont {Aguado}}]{Prada2012Transport}%
  \BibitemOpen
  \bibfield  {author} {\bibinfo {author} {\bibfnamefont {E.}~\bibnamefont
  {Prada}}, \bibinfo {author} {\bibfnamefont {P.}~\bibnamefont {San-Jose}}, \
  and\ \bibinfo {author} {\bibfnamefont {R.}~\bibnamefont {Aguado}},\ }\href
  {\doibase 10.1103/PhysRevB.86.180503} {\bibfield  {journal} {\bibinfo
  {journal} {Phys. Rev. B}\ }\textbf {\bibinfo {volume} {86}},\ \bibinfo
  {pages} {180503} (\bibinfo {year} {2012})}\BibitemShut {NoStop}%
\bibitem [{\citenamefont {Liu}\ \emph {et~al.}(2017{\natexlab{a}})\citenamefont
  {Liu}, \citenamefont {Sau}, \citenamefont {Stanescu},\ and\ \citenamefont
  {Das~Sarma}}]{Liu2017Andreev}%
  \BibitemOpen
  \bibfield  {author} {\bibinfo {author} {\bibfnamefont {C.-X.}\ \bibnamefont
  {Liu}}, \bibinfo {author} {\bibfnamefont {J.~D.}\ \bibnamefont {Sau}},
  \bibinfo {author} {\bibfnamefont {T.~D.}\ \bibnamefont {Stanescu}}, \ and\
  \bibinfo {author} {\bibfnamefont {S.}~\bibnamefont {Das~Sarma}},\ }\href
  {\doibase 10.1103/PhysRevB.96.075161} {\bibfield  {journal} {\bibinfo
  {journal} {Phys. Rev. B}\ }\textbf {\bibinfo {volume} {96}},\ \bibinfo
  {pages} {075161} (\bibinfo {year} {2017}{\natexlab{a}})}\BibitemShut
  {NoStop}%
\bibitem [{\citenamefont {Chiu}\ \emph {et~al.}(2017)\citenamefont {Chiu},
  \citenamefont {Sau},\ and\ \citenamefont {Das~Sarma}}]{Chiu2017Conductance}%
  \BibitemOpen
  \bibfield  {author} {\bibinfo {author} {\bibfnamefont {C.-K.}\ \bibnamefont
  {Chiu}}, \bibinfo {author} {\bibfnamefont {J.~D.}\ \bibnamefont {Sau}}, \
  and\ \bibinfo {author} {\bibfnamefont {S.}~\bibnamefont {Das~Sarma}},\ }\href
  {\doibase 10.1103/PhysRevB.96.054504} {\bibfield  {journal} {\bibinfo
  {journal} {Phys. Rev. B}\ }\textbf {\bibinfo {volume} {96}},\ \bibinfo
  {pages} {054504} (\bibinfo {year} {2017})}\BibitemShut {NoStop}%
\bibitem [{\citenamefont {Setiawan}\ \emph {et~al.}(2017)\citenamefont
  {Setiawan}, \citenamefont {Liu}, \citenamefont {Sau},\ and\ \citenamefont
  {Das~Sarma}}]{Setiawan2017Electron}%
  \BibitemOpen
  \bibfield  {author} {\bibinfo {author} {\bibfnamefont {F.}~\bibnamefont
  {Setiawan}}, \bibinfo {author} {\bibfnamefont {C.-X.}\ \bibnamefont {Liu}},
  \bibinfo {author} {\bibfnamefont {J.~D.}\ \bibnamefont {Sau}}, \ and\
  \bibinfo {author} {\bibfnamefont {S.}~\bibnamefont {Das~Sarma}},\ }\href
  {\doibase 10.1103/PhysRevB.96.184520} {\bibfield  {journal} {\bibinfo
  {journal} {Phys. Rev. B}\ }\textbf {\bibinfo {volume} {96}},\ \bibinfo
  {pages} {184520} (\bibinfo {year} {2017})}\BibitemShut {NoStop}%
\bibitem [{\citenamefont {Moore}\ \emph
  {et~al.}(2018{\natexlab{a}})\citenamefont {Moore}, \citenamefont {Stanescu},\
  and\ \citenamefont {Tewari}}]{Moore2018a}%
  \BibitemOpen
  \bibfield  {author} {\bibinfo {author} {\bibfnamefont {C.}~\bibnamefont
  {Moore}}, \bibinfo {author} {\bibfnamefont {T.~D.}\ \bibnamefont {Stanescu}},
  \ and\ \bibinfo {author} {\bibfnamefont {S.}~\bibnamefont {Tewari}},\ }\href
  {\doibase 10.1103/PhysRevB.97.165302} {\bibfield  {journal} {\bibinfo
  {journal} {Phys. Rev. B}\ }\textbf {\bibinfo {volume} {97}},\ \bibinfo
  {pages} {165302} (\bibinfo {year} {2018}{\natexlab{a}})}\BibitemShut
  {NoStop}%
\bibitem [{\citenamefont {Liu}\ \emph {et~al.}(2018{\natexlab{a}})\citenamefont
  {Liu}, \citenamefont {Sau},\ and\ \citenamefont
  {Das~Sarma}}]{Liu2018Distinguishing}%
  \BibitemOpen
  \bibfield  {author} {\bibinfo {author} {\bibfnamefont {C.-X.}\ \bibnamefont
  {Liu}}, \bibinfo {author} {\bibfnamefont {J.~D.}\ \bibnamefont {Sau}}, \ and\
  \bibinfo {author} {\bibfnamefont {S.}~\bibnamefont {Das~Sarma}},\ }\href
  {\doibase 10.1103/PhysRevB.97.214502} {\bibfield  {journal} {\bibinfo
  {journal} {Phys. Rev. B}\ }\textbf {\bibinfo {volume} {97}},\ \bibinfo
  {pages} {214502} (\bibinfo {year} {2018}{\natexlab{a}})}\BibitemShut
  {NoStop}%
\bibitem [{\citenamefont {Fleckenstein}\ \emph {et~al.}(2018)\citenamefont
  {Fleckenstein}, \citenamefont {Dom\'{\i}nguez}, \citenamefont
  {Traverso~Ziani},\ and\ \citenamefont
  {Trauzettel}}]{Fleckenstein2018Decaying}%
  \BibitemOpen
  \bibfield  {author} {\bibinfo {author} {\bibfnamefont {C.}~\bibnamefont
  {Fleckenstein}}, \bibinfo {author} {\bibfnamefont {F.}~\bibnamefont
  {Dom\'{\i}nguez}}, \bibinfo {author} {\bibfnamefont {N.}~\bibnamefont
  {Traverso~Ziani}}, \ and\ \bibinfo {author} {\bibfnamefont {B.}~\bibnamefont
  {Trauzettel}},\ }\href {\doibase 10.1103/PhysRevB.97.155425} {\bibfield
  {journal} {\bibinfo  {journal} {Phys. Rev. B}\ }\textbf {\bibinfo {volume}
  {97}},\ \bibinfo {pages} {155425} (\bibinfo {year} {2018})}\BibitemShut
  {NoStop}%
\bibitem [{\citenamefont {Moore}\ \emph
  {et~al.}(2018{\natexlab{b}})\citenamefont {Moore}, \citenamefont {Zeng},
  \citenamefont {Stanescu},\ and\ \citenamefont {Tewari}}]{Moore2018b}%
  \BibitemOpen
  \bibfield  {author} {\bibinfo {author} {\bibfnamefont {C.}~\bibnamefont
  {Moore}}, \bibinfo {author} {\bibfnamefont {C.}~\bibnamefont {Zeng}},
  \bibinfo {author} {\bibfnamefont {T.~D.}\ \bibnamefont {Stanescu}}, \ and\
  \bibinfo {author} {\bibfnamefont {S.}~\bibnamefont {Tewari}},\ }\href
  {\doibase 10.1103/PhysRevB.98.155314} {\bibfield  {journal} {\bibinfo
  {journal} {Phys. Rev. B}\ }\textbf {\bibinfo {volume} {98}},\ \bibinfo
  {pages} {155314} (\bibinfo {year} {2018}{\natexlab{b}})}\BibitemShut
  {NoStop}%
\bibitem [{\citenamefont {Vuik}\ \emph {et~al.}(2018)\citenamefont {Vuik},
  \citenamefont {Nijholt}, \citenamefont {Akhmerov},\ and\ \citenamefont
  {Wimmer}}]{Vuik2018Reproducing}%
  \BibitemOpen
  \bibfield  {author} {\bibinfo {author} {\bibfnamefont {A.}~\bibnamefont
  {Vuik}}, \bibinfo {author} {\bibfnamefont {B.}~\bibnamefont {Nijholt}},
  \bibinfo {author} {\bibfnamefont {A.}~\bibnamefont {Akhmerov}}, \ and\
  \bibinfo {author} {\bibfnamefont {M.}~\bibnamefont {Wimmer}},\ }\href
  {https://arxiv.org/abs/1806.02801} {\bibfield  {journal} {\bibinfo  {journal}
  {arXiv:1806.02801}\ } (\bibinfo {year} {2018})}\BibitemShut {NoStop}%
\bibitem [{\citenamefont {Stanescu}\ \emph {et~al.}(2011)\citenamefont
  {Stanescu}, \citenamefont {Lutchyn},\ and\ \citenamefont
  {Das~Sarma}}]{Stanescu2011Majorana}%
  \BibitemOpen
  \bibfield  {author} {\bibinfo {author} {\bibfnamefont {T.~D.}\ \bibnamefont
  {Stanescu}}, \bibinfo {author} {\bibfnamefont {R.~M.}\ \bibnamefont
  {Lutchyn}}, \ and\ \bibinfo {author} {\bibfnamefont {S.}~\bibnamefont
  {Das~Sarma}},\ }\href {\doibase 10.1103/PhysRevB.84.144522} {\bibfield
  {journal} {\bibinfo  {journal} {Phys. Rev. B}\ }\textbf {\bibinfo {volume}
  {84}},\ \bibinfo {pages} {144522} (\bibinfo {year} {2011})}\BibitemShut
  {NoStop}%
\bibitem [{\citenamefont {Liu}\ \emph {et~al.}(2017{\natexlab{b}})\citenamefont
  {Liu}, \citenamefont {Sau},\ and\ \citenamefont {Das~Sarma}}]{Liu2017Role}%
  \BibitemOpen
  \bibfield  {author} {\bibinfo {author} {\bibfnamefont {C.-X.}\ \bibnamefont
  {Liu}}, \bibinfo {author} {\bibfnamefont {J.~D.}\ \bibnamefont {Sau}}, \ and\
  \bibinfo {author} {\bibfnamefont {S.}~\bibnamefont {Das~Sarma}},\ }\href
  {\doibase 10.1103/PhysRevB.95.054502} {\bibfield  {journal} {\bibinfo
  {journal} {Phys. Rev. B}\ }\textbf {\bibinfo {volume} {95}},\ \bibinfo
  {pages} {054502} (\bibinfo {year} {2017}{\natexlab{b}})}\BibitemShut
  {NoStop}%
\bibitem [{\citenamefont {Chiu}\ \emph {et~al.}(2018)\citenamefont {Chiu},
  \citenamefont {Sau},\ and\ \citenamefont {Das~Sarma}}]{Chiu2017Interference}%
  \BibitemOpen
  \bibfield  {author} {\bibinfo {author} {\bibfnamefont {C.-K.}\ \bibnamefont
  {Chiu}}, \bibinfo {author} {\bibfnamefont {J.~D.}\ \bibnamefont {Sau}}, \
  and\ \bibinfo {author} {\bibfnamefont {S.}~\bibnamefont {Das~Sarma}},\ }\href
  {\doibase 10.1103/PhysRevB.97.035310} {\bibfield  {journal} {\bibinfo
  {journal} {Phys. Rev. B}\ }\textbf {\bibinfo {volume} {97}},\ \bibinfo
  {pages} {035310} (\bibinfo {year} {2018})}\BibitemShut {NoStop}%
\bibitem [{\citenamefont {Chiu}\ and\ \citenamefont
  {Sarma}(2018)}]{Chiu2018Fractional}%
  \BibitemOpen
  \bibfield  {author} {\bibinfo {author} {\bibfnamefont {C.-K.}\ \bibnamefont
  {Chiu}}\ and\ \bibinfo {author} {\bibfnamefont {S.~D.}\ \bibnamefont
  {Sarma}},\ }\href {https://arxiv.org/abs/1806.02224} {\bibfield  {journal}
  {\bibinfo  {journal} {arXiv:1806.02224}\ } (\bibinfo {year}
  {2018})}\BibitemShut {NoStop}%
\bibitem [{\citenamefont {Liu}\ \emph {et~al.}(2018{\natexlab{b}})\citenamefont
  {Liu}, \citenamefont {Cole},\ and\ \citenamefont {Sau}}]{Liu2018Measuring}%
  \BibitemOpen
  \bibfield  {author} {\bibinfo {author} {\bibfnamefont {C.-X.}\ \bibnamefont
  {Liu}}, \bibinfo {author} {\bibfnamefont {W.~S.}\ \bibnamefont {Cole}}, \
  and\ \bibinfo {author} {\bibfnamefont {J.~D.}\ \bibnamefont {Sau}},\ }\href
  {https://arxiv.org/abs/1803.01872} {\bibfield  {journal} {\bibinfo  {journal}
  {arXiv:1803.01872}\ } (\bibinfo {year} {2018}{\natexlab{b}})}\BibitemShut
  {NoStop}%
\bibitem [{\citenamefont {Bommer}\ \emph {et~al.}(2018)\citenamefont {Bommer},
  \citenamefont {Zhang}, \citenamefont {G{\"u}l}, \citenamefont {Nijholt},
  \citenamefont {Wimmer}, \citenamefont {Rybakov}, \citenamefont {Garaud},
  \citenamefont {Rodic}, \citenamefont {Babaev}, \citenamefont {Troyer} \emph
  {et~al.}}]{Bommer2018SpinOrbit}%
  \BibitemOpen
  \bibfield  {author} {\bibinfo {author} {\bibfnamefont {J.~D.}\ \bibnamefont
  {Bommer}}, \bibinfo {author} {\bibfnamefont {H.}~\bibnamefont {Zhang}},
  \bibinfo {author} {\bibfnamefont {{\"O}.}~\bibnamefont {G{\"u}l}}, \bibinfo
  {author} {\bibfnamefont {B.}~\bibnamefont {Nijholt}}, \bibinfo {author}
  {\bibfnamefont {M.}~\bibnamefont {Wimmer}}, \bibinfo {author} {\bibfnamefont
  {F.~N.}\ \bibnamefont {Rybakov}}, \bibinfo {author} {\bibfnamefont
  {J.}~\bibnamefont {Garaud}}, \bibinfo {author} {\bibfnamefont
  {D.}~\bibnamefont {Rodic}}, \bibinfo {author} {\bibfnamefont
  {E.}~\bibnamefont {Babaev}}, \bibinfo {author} {\bibfnamefont
  {M.}~\bibnamefont {Troyer}},  \emph {et~al.},\ }\href
  {https://arxiv.org/abs/1807.01940} {\bibfield  {journal} {\bibinfo  {journal}
  {arXiv:1807.01940}\ } (\bibinfo {year} {2018})}\BibitemShut {NoStop}%
\bibitem [{\citenamefont {Beenakker}(1992)}]{Beenakker1992Quantum}%
  \BibitemOpen
  \bibfield  {author} {\bibinfo {author} {\bibfnamefont {C.~W.~J.}\
  \bibnamefont {Beenakker}},\ }\href {\doibase 10.1103/PhysRevB.46.12841}
  {\bibfield  {journal} {\bibinfo  {journal} {Phys. Rev. B}\ }\textbf {\bibinfo
  {volume} {46}},\ \bibinfo {pages} {12841} (\bibinfo {year}
  {1992})}\BibitemShut {NoStop}%
\bibitem [{\citenamefont {Blonder}\ \emph {et~al.}(1982)\citenamefont
  {Blonder}, \citenamefont {Tinkham},\ and\ \citenamefont
  {Klapwijk}}]{Blonder1982Transition}%
  \BibitemOpen
  \bibfield  {author} {\bibinfo {author} {\bibfnamefont {G.~E.}\ \bibnamefont
  {Blonder}}, \bibinfo {author} {\bibfnamefont {M.}~\bibnamefont {Tinkham}}, \
  and\ \bibinfo {author} {\bibfnamefont {T.~M.}\ \bibnamefont {Klapwijk}},\
  }\href {\doibase 10.1103/PhysRevB.25.4515} {\bibfield  {journal} {\bibinfo
  {journal} {Phys. Rev. B}\ }\textbf {\bibinfo {volume} {25}},\ \bibinfo
  {pages} {4515} (\bibinfo {year} {1982})}\BibitemShut {NoStop}%
\bibitem [{\citenamefont {Groth}\ \emph {et~al.}(2014)\citenamefont {Groth},
  \citenamefont {Wimmer}, \citenamefont {Akhmerov},\ and\ \citenamefont
  {Waintal}}]{kwant}%
  \BibitemOpen
  \bibfield  {author} {\bibinfo {author} {\bibfnamefont {C.~W.}\ \bibnamefont
  {Groth}}, \bibinfo {author} {\bibfnamefont {M.}~\bibnamefont {Wimmer}},
  \bibinfo {author} {\bibfnamefont {A.~R.}\ \bibnamefont {Akhmerov}}, \ and\
  \bibinfo {author} {\bibfnamefont {X.}~\bibnamefont {Waintal}},\ }\href@noop
  {} {\bibfield  {journal} {\bibinfo  {journal} {New Journal of Physics}\
  }\textbf {\bibinfo {volume} {16}},\ \bibinfo {pages} {063065} (\bibinfo
  {year} {2014})}\BibitemShut {NoStop}%
\bibitem [{\citenamefont {Lin}\ \emph {et~al.}(2012)\citenamefont {Lin},
  \citenamefont {Sau},\ and\ \citenamefont {Das~Sarma}}]{Lin2012Zero}%
  \BibitemOpen
  \bibfield  {author} {\bibinfo {author} {\bibfnamefont {C.-H.}\ \bibnamefont
  {Lin}}, \bibinfo {author} {\bibfnamefont {J.~D.}\ \bibnamefont {Sau}}, \ and\
  \bibinfo {author} {\bibfnamefont {S.}~\bibnamefont {Das~Sarma}},\ }\href
  {\doibase 10.1103/PhysRevB.86.224511} {\bibfield  {journal} {\bibinfo
  {journal} {Phys. Rev. B}\ }\textbf {\bibinfo {volume} {86}},\ \bibinfo
  {pages} {224511} (\bibinfo {year} {2012})}\BibitemShut {NoStop}%
\bibitem [{\citenamefont {Liu}\ \emph {et~al.}(2017{\natexlab{c}})\citenamefont
  {Liu}, \citenamefont {Setiawan}, \citenamefont {Sau},\ and\ \citenamefont
  {Das~Sarma}}]{Liu2017Phenomenology}%
  \BibitemOpen
  \bibfield  {author} {\bibinfo {author} {\bibfnamefont {C.-X.}\ \bibnamefont
  {Liu}}, \bibinfo {author} {\bibfnamefont {F.}~\bibnamefont {Setiawan}},
  \bibinfo {author} {\bibfnamefont {J.~D.}\ \bibnamefont {Sau}}, \ and\
  \bibinfo {author} {\bibfnamefont {S.}~\bibnamefont {Das~Sarma}},\ }\href
  {\doibase 10.1103/PhysRevB.96.054520} {\bibfield  {journal} {\bibinfo
  {journal} {Phys. Rev. B}\ }\textbf {\bibinfo {volume} {96}},\ \bibinfo
  {pages} {054520} (\bibinfo {year} {2017}{\natexlab{c}})}\BibitemShut
  {NoStop}%
\bibitem [{\citenamefont {de~Moor}\ \emph {et~al.}(2018)\citenamefont
  {de~Moor}, \citenamefont {Bommer}, \citenamefont {Xu}, \citenamefont
  {Winkler}, \citenamefont {Antipov}, \citenamefont {Bargerbos}, \citenamefont
  {Wang}, \citenamefont {van Loo}, \citenamefont {het Veld}, \citenamefont
  {Gazibegovic}, \citenamefont {Car}, \citenamefont {Logan}, \citenamefont
  {Pendharkar}, \citenamefont {Lee}, \citenamefont {Bakkers}, \citenamefont
  {Palmstr{\o}m}, \citenamefont {Lutchyn}, \citenamefont {Kouwenhoven},\ and\
  \citenamefont {Zhang}}]{deMoor2018Electric}%
  \BibitemOpen
  \bibfield  {author} {\bibinfo {author} {\bibfnamefont {M.~W.~A.}\
  \bibnamefont {de~Moor}}, \bibinfo {author} {\bibfnamefont {J.~D.~S.}\
  \bibnamefont {Bommer}}, \bibinfo {author} {\bibfnamefont {D.}~\bibnamefont
  {Xu}}, \bibinfo {author} {\bibfnamefont {G.~W.}\ \bibnamefont {Winkler}},
  \bibinfo {author} {\bibfnamefont {A.~E.}\ \bibnamefont {Antipov}}, \bibinfo
  {author} {\bibfnamefont {A.}~\bibnamefont {Bargerbos}}, \bibinfo {author}
  {\bibfnamefont {G.}~\bibnamefont {Wang}}, \bibinfo {author} {\bibfnamefont
  {N.}~\bibnamefont {van Loo}}, \bibinfo {author} {\bibfnamefont {R.~L. M.~O.}\
  \bibnamefont {het Veld}}, \bibinfo {author} {\bibfnamefont {S.}~\bibnamefont
  {Gazibegovic}}, \bibinfo {author} {\bibfnamefont {D.}~\bibnamefont {Car}},
  \bibinfo {author} {\bibfnamefont {J.~A.}\ \bibnamefont {Logan}}, \bibinfo
  {author} {\bibfnamefont {M.}~\bibnamefont {Pendharkar}}, \bibinfo {author}
  {\bibfnamefont {J.~S.}\ \bibnamefont {Lee}}, \bibinfo {author} {\bibfnamefont
  {E.~P. A.~M.}\ \bibnamefont {Bakkers}}, \bibinfo {author} {\bibfnamefont
  {C.~J.}\ \bibnamefont {Palmstr{\o}m}}, \bibinfo {author} {\bibfnamefont
  {R.~M.}\ \bibnamefont {Lutchyn}}, \bibinfo {author} {\bibfnamefont {L.~P.}\
  \bibnamefont {Kouwenhoven}}, \ and\ \bibinfo {author} {\bibfnamefont
  {H.}~\bibnamefont {Zhang}},\ }\href
  {http://stacks.iop.org/1367-2630/20/i=10/a=103049} {\bibfield  {journal}
  {\bibinfo  {journal} {New Journal of Physics}\ }\textbf {\bibinfo {volume}
  {20}},\ \bibinfo {pages} {103049} (\bibinfo {year} {2018})}\BibitemShut
  {NoStop}%
\bibitem [{\citenamefont {Woods}\ \emph {et~al.}(2018)\citenamefont {Woods},
  \citenamefont {Stanescu},\ and\ \citenamefont
  {Das~Sarma}}]{Woods2018Effective}%
  \BibitemOpen
  \bibfield  {author} {\bibinfo {author} {\bibfnamefont {B.~D.}\ \bibnamefont
  {Woods}}, \bibinfo {author} {\bibfnamefont {T.~D.}\ \bibnamefont {Stanescu}},
  \ and\ \bibinfo {author} {\bibfnamefont {S.}~\bibnamefont {Das~Sarma}},\
  }\href {\doibase 10.1103/PhysRevB.98.035428} {\bibfield  {journal} {\bibinfo
  {journal} {Phys. Rev. B}\ }\textbf {\bibinfo {volume} {98}},\ \bibinfo
  {pages} {035428} (\bibinfo {year} {2018})}\BibitemShut {NoStop}%
\bibitem [{\citenamefont {Antipov}\ \emph {et~al.}(2018)\citenamefont
  {Antipov}, \citenamefont {Bargerbos}, \citenamefont {Winkler}, \citenamefont
  {Bauer}, \citenamefont {Rossi},\ and\ \citenamefont
  {Lutchyn}}]{Antipov2018Effects}%
  \BibitemOpen
  \bibfield  {author} {\bibinfo {author} {\bibfnamefont {A.~E.}\ \bibnamefont
  {Antipov}}, \bibinfo {author} {\bibfnamefont {A.}~\bibnamefont {Bargerbos}},
  \bibinfo {author} {\bibfnamefont {G.~W.}\ \bibnamefont {Winkler}}, \bibinfo
  {author} {\bibfnamefont {B.}~\bibnamefont {Bauer}}, \bibinfo {author}
  {\bibfnamefont {E.}~\bibnamefont {Rossi}}, \ and\ \bibinfo {author}
  {\bibfnamefont {R.~M.}\ \bibnamefont {Lutchyn}},\ }\href {\doibase
  10.1103/PhysRevX.8.031041} {\bibfield  {journal} {\bibinfo  {journal} {Phys.
  Rev. X}\ }\textbf {\bibinfo {volume} {8}},\ \bibinfo {pages} {031041}
  (\bibinfo {year} {2018})}\BibitemShut {NoStop}%
\bibitem [{\citenamefont {Mikkelsen}\ \emph {et~al.}(2018)\citenamefont
  {Mikkelsen}, \citenamefont {Kotetes}, \citenamefont {Krogstrup},\ and\
  \citenamefont {Flensberg}}]{Mikkelsen2018Hybridization}%
  \BibitemOpen
  \bibfield  {author} {\bibinfo {author} {\bibfnamefont {A.~E.~G.}\
  \bibnamefont {Mikkelsen}}, \bibinfo {author} {\bibfnamefont {P.}~\bibnamefont
  {Kotetes}}, \bibinfo {author} {\bibfnamefont {P.}~\bibnamefont {Krogstrup}},
  \ and\ \bibinfo {author} {\bibfnamefont {K.}~\bibnamefont {Flensberg}},\
  }\href {\doibase 10.1103/PhysRevX.8.031040} {\bibfield  {journal} {\bibinfo
  {journal} {Phys. Rev. X}\ }\textbf {\bibinfo {volume} {8}},\ \bibinfo {pages}
  {031040} (\bibinfo {year} {2018})}\BibitemShut {NoStop}%
\bibitem [{\citenamefont {Huang}\ \emph {et~al.}(2018)\citenamefont {Huang},
  \citenamefont {Pan}, \citenamefont {Liu}, \citenamefont {Sau}, \citenamefont
  {Stanescu},\ and\ \citenamefont {Das~Sarma}}]{Huang2018Metamorphosis}%
  \BibitemOpen
  \bibfield  {author} {\bibinfo {author} {\bibfnamefont {Y.}~\bibnamefont
  {Huang}}, \bibinfo {author} {\bibfnamefont {H.}~\bibnamefont {Pan}}, \bibinfo
  {author} {\bibfnamefont {C.-X.}\ \bibnamefont {Liu}}, \bibinfo {author}
  {\bibfnamefont {J.~D.}\ \bibnamefont {Sau}}, \bibinfo {author} {\bibfnamefont
  {T.~D.}\ \bibnamefont {Stanescu}}, \ and\ \bibinfo {author} {\bibfnamefont
  {S.}~\bibnamefont {Das~Sarma}},\ }\href {\doibase 10.1103/PhysRevB.98.144511}
  {\bibfield  {journal} {\bibinfo  {journal} {Phys. Rev. B}\ }\textbf {\bibinfo
  {volume} {98}},\ \bibinfo {pages} {144511} (\bibinfo {year}
  {2018})}\BibitemShut {NoStop}%
\bibitem [{\citenamefont {Stanescu}\ \emph {et~al.}(2010)\citenamefont
  {Stanescu}, \citenamefont {Sau}, \citenamefont {Lutchyn},\ and\ \citenamefont
  {Das~Sarma}}]{Stanescu2010Proximity}%
  \BibitemOpen
  \bibfield  {author} {\bibinfo {author} {\bibfnamefont {T.~D.}\ \bibnamefont
  {Stanescu}}, \bibinfo {author} {\bibfnamefont {J.~D.}\ \bibnamefont {Sau}},
  \bibinfo {author} {\bibfnamefont {R.~M.}\ \bibnamefont {Lutchyn}}, \ and\
  \bibinfo {author} {\bibfnamefont {S.}~\bibnamefont {Das~Sarma}},\ }\href
  {\doibase 10.1103/PhysRevB.81.241310} {\bibfield  {journal} {\bibinfo
  {journal} {Phys. Rev. B}\ }\textbf {\bibinfo {volume} {81}},\ \bibinfo
  {pages} {241310} (\bibinfo {year} {2010})}\BibitemShut {NoStop}%
\bibitem [{\citenamefont {Stanescu}\ and\ \citenamefont
  {Das~Sarma}(2017)}]{Stanescu2017Proximity}%
  \BibitemOpen
  \bibfield  {author} {\bibinfo {author} {\bibfnamefont {T.~D.}\ \bibnamefont
  {Stanescu}}\ and\ \bibinfo {author} {\bibfnamefont {S.}~\bibnamefont
  {Das~Sarma}},\ }\href {\doibase 10.1103/PhysRevB.96.014510} {\bibfield
  {journal} {\bibinfo  {journal} {Phys. Rev. B}\ }\textbf {\bibinfo {volume}
  {96}},\ \bibinfo {pages} {014510} (\bibinfo {year} {2017})}\BibitemShut
  {NoStop}%
\bibitem [{\citenamefont {Reeg}\ and\ \citenamefont
  {Maslov}(2017)}]{Reeg2017Transport}%
  \BibitemOpen
  \bibfield  {author} {\bibinfo {author} {\bibfnamefont {C.}~\bibnamefont
  {Reeg}}\ and\ \bibinfo {author} {\bibfnamefont {D.~L.}\ \bibnamefont
  {Maslov}},\ }\href {\doibase 10.1103/PhysRevB.95.205439} {\bibfield
  {journal} {\bibinfo  {journal} {Phys. Rev. B}\ }\textbf {\bibinfo {volume}
  {95}},\ \bibinfo {pages} {205439} (\bibinfo {year} {2017})}\BibitemShut
  {NoStop}%
\bibitem [{\citenamefont {Stenger}\ and\ \citenamefont
  {Stanescu}(2017)}]{Stenger2017Transport}%
  \BibitemOpen
  \bibfield  {author} {\bibinfo {author} {\bibfnamefont {J.}~\bibnamefont
  {Stenger}}\ and\ \bibinfo {author} {\bibfnamefont {T.~D.}\ \bibnamefont
  {Stanescu}},\ }\href {\doibase 10.1103/PhysRevB.96.214516} {\bibfield
  {journal} {\bibinfo  {journal} {Phys. Rev. B}\ }\textbf {\bibinfo {volume}
  {96}},\ \bibinfo {pages} {214516} (\bibinfo {year} {2017})}\BibitemShut
  {NoStop}%
\bibitem [{\citenamefont {Danon}\ \emph {et~al.}(2017)\citenamefont {Danon},
  \citenamefont {Hansen},\ and\ \citenamefont
  {Flensberg}}]{Danon2017Conductance}%
  \BibitemOpen
  \bibfield  {author} {\bibinfo {author} {\bibfnamefont {J.}~\bibnamefont
  {Danon}}, \bibinfo {author} {\bibfnamefont {E.~B.}\ \bibnamefont {Hansen}}, \
  and\ \bibinfo {author} {\bibfnamefont {K.}~\bibnamefont {Flensberg}},\ }\href
  {\doibase 10.1103/PhysRevB.96.125420} {\bibfield  {journal} {\bibinfo
  {journal} {Phys. Rev. B}\ }\textbf {\bibinfo {volume} {96}},\ \bibinfo
  {pages} {125420} (\bibinfo {year} {2017})}\BibitemShut {NoStop}%
\bibitem [{\citenamefont {Martin}\ and\ \citenamefont
  {Mozyrsky}(2014)}]{Martin2014Noneq}%
  \BibitemOpen
  \bibfield  {author} {\bibinfo {author} {\bibfnamefont {I.}~\bibnamefont
  {Martin}}\ and\ \bibinfo {author} {\bibfnamefont {D.}~\bibnamefont
  {Mozyrsky}},\ }\href {\doibase 10.1103/PhysRevB.90.100508} {\bibfield
  {journal} {\bibinfo  {journal} {Phys. Rev. B}\ }\textbf {\bibinfo {volume}
  {90}},\ \bibinfo {pages} {100508} (\bibinfo {year} {2014})}\BibitemShut
  {NoStop}%
\bibitem [{\citenamefont {Stanescu}\ and\ \citenamefont
  {Tewari}(2018)}]{Stanescu2018Illustrated}%
  \BibitemOpen
  \bibfield  {author} {\bibinfo {author} {\bibfnamefont {T.~D.}\ \bibnamefont
  {Stanescu}}\ and\ \bibinfo {author} {\bibfnamefont {S.}~\bibnamefont
  {Tewari}},\ }\href {https://arxiv.org/abs/1811.02557} {\bibfield  {journal}
  {\bibinfo  {journal} {arXiv:1811.02557}\ } (\bibinfo {year}
  {2018})}\BibitemShut {NoStop}%
\end{thebibliography}%

\end{document}